\documentclass[sn-aps]{sn-jnl}

\jyear{2022}%

\begin{document}

\title[Understanding the relativistic overdensity of galaxy surveys]{Understanding the relativistic overdensity of galaxy surveys}

%
%
%
%
%

\author*{\fnm{Didam G.A.} \sur{Duniya}}\email{duniyaa@biust.ac.bw}

\affil*[1]{\orgdiv{Department of Physics \& Astronomy}, \orgname{Botswana International University of Science and Technology}, \orgaddress{\city{Palapye}, \country{Botswana}}}

\affil[2]{\orgdiv{Department of Mathematics \& Applied Mathematics},\\ \orgname{University of Cape Town}, \orgaddress{\city{Cape Town}, \postcode{7701}, \country{South Africa}}}

\affil[3]{\orgname{African Institute for Mathematical Sciences (AIMS)},\\ \orgaddress{\city{Cape Town} \postcode{7945}, \country{South Africa}}}

\affil[4]{\orgdiv{Department of Physics \& Astronomy}, \orgname{University of the Western Cape}, \orgaddress{\city{Cape Town} \postcode{7535}, \country{South Africa}}}

\abstract{The main goal of galaxy surveys is to map the distribution of the galaxies, for the purpose of understanding the properties of this distribution and its implications for the content and the evolution of the universe. However, in order to realise the full potential of these surveys, we need to ensure that we are using the correct analysis: a relativistic analysis, which has been widely studied recently. In this work, the known relativistic overdensity of galaxy surveys is re-examined. Unlike in previous works, a consistent approach for incorporating both the relativistic number-count overdensity and the relativistic cosmic magnification overdensity in the total observed overdensity of a generic survey, is presented. Since in practice, analyses are often done for specific sample types (flux-limited or volume-limited) the approach in this work allows for the total observed overdensity to be easily reduced to either of the individual overdensities by applying the same limiting conditions as for extracting the corresponding data samples. This is not obvious, or in some cases not possible, with the results of previous works. Thus, the calculations in this work serve to generalise the expression for the total observed overdensity. However, care must be taken to apply it appropriately: the type of the data sample in consideration needs to be taken into account.}




\maketitle

\section{Introduction}\label{sec:intro}
In recent years, relativistic effects \citep{Yoo:2009au, Yoo:2010ni, Bonvin:2011bg, Challinor:2011bk, Jeong:2011as, Duniya:2015ths, Bonvin:2014owa, LopezHonorez:2011cy, Alonso:2015uua, Renk:2016olm, Durrer:2016jzq, Duniya:2016gcf, Duniya:2013eta, Montanari:2015rga, Bartolo:2010ec, Baldauf:2011bh, Yoo:2011zc, Schmidt:2012ne, Jeong:2012nu, Schmidt:2012nw, Bertacca:2012tp, Maartens:2012rh, Flender:2012nq, Hall:2012wd, Lombriser:2013aj, Yoo:2013tc, Jeong:2013psa, Raccanelli:2013dza, DiDio:2013bqa, Bonvin:2013ogt, Raccanelli:2013gja, Bacon:2014uja, Jeong:2014ufa, Villa:2014foa, Yoo:2014kpa, Camera:2014bwa, Camera:2014sba, Duniya:2015nva, Kehagias:2015tda, Duniya:2015dpa, Raccanelli:2015vla, Bartolo:2015qva, Alonso:2015sfa, Yoo:2015uxa, Irsic:2015nla, DiDio:2015bua, Bonvin:2015kuc, Gaztanaga:2015jrs, Raccanelli:2016avd, Duniya:2019mpr} have been vastly studied. A number of the studies in the literature have shown how to calculate the observed clustering overdensity of galaxies in a relativistic context, i.e.~the relativistic galaxy overdensity~\citep{Yoo:2009au, Yoo:2010ni, Bonvin:2011bg, Challinor:2011bk, Jeong:2011as, LopezHonorez:2011cy, Bonvin:2014owa, Duniya:2015ths, Alonso:2015uua, Renk:2016olm, Durrer:2016jzq}, revealing relativistic corrections to the simple ``Newtonian'' calculation that is standard for low redshift, small area surveys. Galaxy surveys with large volume coverage---spanning scales nearly and larger than the Hubble radius, reaching high redshifts---require relativistic corrections to the observed overdensity. Basically, these corrections are owing to the large-scale effects of (i) the gravitational potential, both local at the observed galaxies and also along the line of sight, and (ii) peculiar velocities, by the apparent motion of the galaxies relative to the observer. Together, these constitute the so-called ``relativistic effects''. 

Current optical surveys only cover low redshifts and sky area with dimensions which are smaller than the Hubble radius; thus, containing negligible relativistic effects. However, upcoming surveys in the optical and the radio/infrared, will extend to high redshifts and encompass large sky area---with dimensions equal and larger than the Hubble radius---providing deeper and richer information on the observable universe but, most importantly, on cosmological scales at the survey redshifts. On these scales, the relativistic effects become significant. These effects can, e.g.~provide a means for a sensitive probe of the nature of dark energy and modified gravity: by comparing the predictions from the relativistic corrections to the observation of the forthcoming surveys.

There are two fundamental issues underlying the relativistic analysis: Firstly, we need to correctly identify the galaxy overdensity that is observed on the past light cone; secondly, we need to account for all the distortions arising from observing on the past light cone, including redshift and volume perturbations (with all relativistic effects included) \citep{Duniya:2015ths}. These relativistic effects appear in the power spectrum of matter in redshift space. However, all of these effects together can suitably be accounted for---simply by using the fact that the physical (observed) number of the galaxies depends on two main quantities: the survey {\it volume} and the apparent {\it flux} of the sources.

In this work we re-trace the relativistic, observed galaxy overdensity (in first order perturbations) using a simple, consistent approach. We recover all previously calculated terms and, still uncover subtle, yet crucial (background) parameters which are not incorporated by previous works. We start by discussing propagation of photons in Sec. \ref{sec:photonprop}. In Sec. \ref{sec:RDP} we re-calculate the observed, relativistic overdensity of galaxy surveys---while revealing the new parameters; in Sec. \ref{sec:FVLS} we give an overview of specific (flux-limited and volume-limited) samples of galaxy surveys and, in Sec. \ref{sec:OAPS} we illustrate some hypothetical implications of the new parameters. We conclude in Sec. \ref{sec:Concl}. 


\section{Propagation of Photons}\label{sec:photonprop} 

Cosmological data collection and analysis is made possible by astronomers' ability to observe photons emitted by cosmic objects. It is therefore important to discuss the behaviour of photons through gravity.

Hereafter, indices of Greek alphabets denote spacetime, with e.g.~$\mu \,{=}\, 0, i$ (with $i \,{=}\, 1,\, 2,\, 3$), where an index of $0$ denotes temporal component and, Roman-numeral indices (e.g.~$i,\, j,\, k,\,l$) denote spatial components.

\subsection{The Geodesic Equation}\label{subsec:geodeqn} 
All bodies which are under the effect of gravity alone are taken to follow paths known as `geodesics'. Consider an arbitrary path $x^\mu$ in spacetime. Then an infinitesimal deformation $\delta{x}^\mu$ about any given point along the path will induce an infinitesimal distance which is measured by the spacetime metric $ds$:
\begin{equation}\label{ds2}
ds^2 = g_{\mu\nu} dx^\mu dx^\nu,
\end{equation}
where $g_{\mu\nu} \,{=}\, g_{\mu\nu}(x^\alpha)$ is the spacetime metric tensor, which is symmetric. If the path is deformed at every point, from the initial point $i$ to the final point $f$, then $x^\mu$ is said to be a geodesic if it satisfies the `stationarity condition'---given by: $\delta(\int^f_i{ds}) \,{=}\, \int^f_i{\delta(ds)} \,{=}\, 0$ \cite{Duniya:2015ths, Durrer:1994zza, Durrer2008, HLP.TKP1994, CarmeliM:1982}. 

By using the stationarity condition, we have the equation for a geodesic:
\begin{equation}\label{geodesic}
n^\mu \nabla_{\mu} n^\nu = 0,
\end{equation}
where $n^\mu \,{=}\, dx^\mu / d\lambda$ is the 4-vector tangent to the geodesic $x^\mu(\lambda)$, with $\lambda$ known as the \emph{affine parameter}~\cite{HLP.TKP1994} (being a parameter that makes a given path obey \eqref{geodesic}). Moreover, the acceleration along a curve $x^\alpha$, is given by $n^\alpha\nabla_{\alpha}n^\beta$ \cite{Durrer:1994zza, Durrer2008}. Thus given \eqref{geodesic}, it follows that there is no acceleration along a geodesic, i.e.~the net external force acting on an object automatically vanishes along a geodesic.

For a given geodesic $x^\mu(\lambda)$, the norm $n^{\mu}n_{\mu}$ is sometimes called the \emph{constant of the motion} of the geodesic, since it remains constant along the geodesic. By suitable parametrization, the constant of the motion can be normalized to $n^{\mu}n_{\mu} \,{=}\, {-}1$ for a time-like geodesic (with $g_{00} \,{<}\, 0$), and $n^{\mu}n_{\mu} \,{=}\, 1$ for a space-like geodesic~\cite{HLP.TKP1994}.

\subsection{The Photon Geodesic Equation}\label{subsec:null}
The infinitesimal distance between any two adjacent points along a given geodesic $x^\mu$ may be given in terms of the tangent 4-vectors by $ds^2 \,{=}\, n^{\mu} n_{\mu}\, d{\lambda}^2$. For $x^\mu$ to be a `null' geodesic, the constant of the motion must vanish: 
\begin{equation}\label{null}
n^{\mu}n_{\mu} = 0, 
\end{equation}
where consequently, $ds \,{=}\, 0$. For a Killing vector $K_\mu$---i.e. where $K_\mu$ satisfies: $K^\nu\nabla_{(\nu} K_{\mu)} \,{=}\, 0$---which is tangent to the geodesic $x^\mu$, if $K_\mu$ is time-like then the energy associated with $n^\mu$ is given by $n^\mu K_{\mu}$ and the norm $n^{\mu}n_{\mu}$, measures the squared rest-mass~\cite{HLP.TKP1994} (i.e.~for $\hbar \,{=}\, c \,{=}\, 1$). 

Geodesics for which $n^{\mu}n_{\mu} \,{\neq}\, 0$ describe the propagation of `massive' (i.e. non-zero rest-mass) test particles in the gravitational field~\cite{CarmeliM:1982}. Thus the null geodesic, as constrained by \eqref{null}, describes the propagation of `massless' (i.e.~zero rest-mass) particles or photons in the gravitational field. Hence the null geodesic is otherwise known as the \emph{photon geodesic}. For the rest of this work, only the photon geodesic will be treated.

\subsection{Perturbing the Photon Geodesic Equation}\label{subsec:pertnull}
Henceforth, we adopt a Friedmann-Robertson-Walker (FRW) universe, with first-order perturbations. We also adopt a conformal transformation on the metric~\eqref{ds2}, which leaves the geodesic equations unchanged, given by 
\begin{eqnarray}\label{conformal-trans}
ds^2 \to d\tilde{s}^2 &=& a^2 ds^2, \\ \label{conformal-ds2}
&=& a(\eta)^2 \Big\lbrace  - (1+2\phi) d{\eta}^2 + 2B_{\mid i}\, d{\eta} dx^i \nonumber\\
&&\hspace{1.1cm} + \left[ (1 - 2\psi)\delta_{ij} +2E_{\mid ij}\right] dx^i dx^j \Big\rbrace ,
\end{eqnarray}
where $\eta$ is the conformal time; $B$, $E$, $\phi$, and $\psi$ are scalar fields, denoting the perturbative degrees of freedom of the metric; with $X_{\mid i} \equiv \nabla_iX$, and $X_{\mid ij} \equiv \nabla_j\nabla_iX$ for a scalar $X$. (See Appendix~\ref{sec:metric} for details on \eqref{conformal-ds2}.) We have $a$ as the `scale factor', which measures the expansion of the universe. Note that until the conformal transformation \eqref{conformal-trans}, $a$ is contained in $g_{\mu\nu}$. Thus, given \eqref{conformal-trans} the metric tensor and the affine parameter transform as follows:
\begin{equation}\label{gmunu-lambda}
\tilde{g}_{\mu\nu} = a^2 g_{\mu\nu}, \quad d\tilde{\lambda} = a^2 d\lambda ,
\end{equation}
and the geodesic tangent 4-vector becomes
\begin{equation}\label{n:Conf}
\tilde{n}^\mu = a^{-2}n^\mu = a^{-2}(1+\delta{n}^0,\, \bar{n}^i +\delta{n}^i),
\end{equation}
where $\bar{n}^0\bar{n}_0 \,{=}\, {-}1$ and $\bar{n}^i \bar{n}_i \,{=}\, 1 \,{=}\, \bar{n}^0$; $\delta{n}^0$ and $\delta{n}^i$ are the temporal and the spatial perturbations, respectively. 

When the nullity condition \eqref{null} holds---as we have adopted for the rest of this work---then the geodesic equation \eqref{geodesic} describes only the propagation of photons. Thus, we can rewrite the photon geodesic equation, given by
\begin{equation}\label{eq:pertgeod1}
\dfrac{dn^\mu}{d\lambda} = -\Gamma^\mu\/_{\alpha\beta}\, n^\alpha n^\beta,
\end{equation}
where $d/d\lambda \,{=}\, n^\mu\partial/\partial{x}^\mu \,{\equiv}\, n^\mu\partial_\mu$ and $\Gamma^\mu\/_{\alpha\beta}$ is the affine connection, given by 
\begin{equation}\label{connection}
\Gamma^\mu\/_{\alpha\beta} = \dfrac{1}{2}g^{\mu\nu}\Big(\partial_\beta g_{\alpha\nu} + \partial_\alpha g_{\nu\beta} - \partial_\nu g_{\alpha\beta}\Big),
\end{equation}
where $g_{\alpha\beta} \,{=}\, \bar{g}_{\alpha\beta} + \delta{g}_{\alpha\beta}$, with its components given by \eqref{conformal-ds2}; the perturbations in $g_{\alpha\beta}$ lead to $\Gamma^\mu\/_{\alpha\beta} \,{=}\, \bar{\Gamma}^\mu\/_{\alpha\beta} + \delta{\Gamma}^\mu\/_{\alpha\beta}$. Given that $\bar{n}^\mu$ obeys \eqref{eq:pertgeod1}, we have 
\begin{equation}\label{eq:pertgeod5}
\dfrac{d\delta{n}^\mu}{d\lambda} = -\bar{g}^{\mu\nu} \dfrac{d}{d\lambda} \left(\bar{n}^\beta\delta{g}_{\nu\beta}\right) + \frac{1}{2}\bar{g}^{\mu\nu} \bar{n}^\alpha\bar{n}^\beta \partial_\nu \delta{g}_{\alpha\beta}, 
\end{equation}
where we used $d\bar{g}_{\mu\nu} / d\lambda \,{=}\, 0$, and hence $\bar{\Gamma}^\mu\/_{\alpha\beta}\bar{n}^\alpha\delta{n}^\beta=0$, with the perturbed affine connection giving
\begin{equation}\label{eq:pertgeod15}\nonumber
\delta{\Gamma}^\mu\/_{\alpha\beta} \bar{n}^\alpha\bar{n}^\beta = \dfrac{1}{2}\bar{g}^{\mu\nu} \bar{n}^\alpha\bar{n}^\beta\left( 2\partial_\alpha \delta{g}_{\nu\beta} - \partial_\nu \delta{g}_{\alpha\beta}\right).
\end{equation}
By integrating the perturbed geodesic equation \eqref{eq:pertgeod5}, we have the perturbed tangent 4-vector, given by
\begin{align} \label{eq:pertgeod19}
\delta{n}^\mu \mid^f_i  = -\Big[\bar{g}^{\mu\nu} \delta{g}_{\nu\beta}\, \bar{n}^\beta\Big]^f_i + \dfrac{1}{2}\bar{g}^{\mu\nu} \int^f_i{ {d\lambda}\, \bar{n}^\alpha\bar{n}^\beta\partial_{\nu}\delta{g}_{\alpha\beta} } .
\end{align}

\subsection{The Photon Displacement}\label{sec:TDV}%
Consider a photon moving in spacetime, from a source $S$ to an observer $O$, along a geodesic ${x}^\mu$. Therefore, we have 
\begin{equation}\label{pertGeod}
\dfrac{d x^\mu}{d\eta} = \dfrac{n^\mu}{n^0} = \bar{n}^\mu +\delta{n}^\mu - \bar{n}^\mu\delta{n}^0, 
\end{equation}
where $x^\mu \,{=}\, \bar{x}^\mu+\delta{x}^\mu$ (similarly for $\eta$), and $n^\mu$ is as given in section \ref{subsec:geodeqn}. By integrating \eqref{pertGeod} from $O$ to $S$:  
\begin{align} \label{posVector}
x^\mu(\eta_S) = -\left(\bar{\eta}_O - \bar{\eta}_S\right)\bar{n}^\mu + \int^{\bar{\eta}_S}_{\bar{\eta}_O}{d\bar{\eta} \left( \delta{n}^\mu - \bar{n}^\mu\delta{n}^0\right)} ,
\end{align}
where $\bar{\eta}_O$ and $\bar{\eta}_S$ are the background times at $O$ and at $S$, respectively; with $\bar{\eta}_O \,{>}\,\bar{\eta}_S$. At lowest order along the photon geodesic we have $d\bar{r} \,{=}\, {-}d\bar{\eta}$, with $r$ being the (radial) comoving distance. (Henceforth, bold-faced quantities denote 3-vectors.) Then by applying the given differential relation, the background part of \eqref{pertGeod} or \eqref{posVector} leads to ${\bf x}(\bar{r}_S) \,{=}\, {-}(\bar{r}_S \,{-}\, \bar{r}_O){\bf n}$: ${\bf n}$ is the unit vector pointing outwards from $O$, with $\bar{r}_S \,{=}\, \bar{r}(\bar{\eta}_S)$ and $\bar{r}_O \,{=}\, \bar{r}(\bar{\eta}_O)$.

Thus at any time instance, the position of the photon relevant to observation is prescribed by the displacement of the observer (from the origin) relative to the instantaneous displacement of the photon (from the origin): 
\begin{equation}\label{displacement}
{\bf x}(\bar{\eta}) = -\left(\bar{r}(\bar{\eta}) - \bar{r}_O\right){\bf n} = -\bar{r}(\bar{\eta}){\bf n},
\end{equation}
where we assume $O$ to be the origin, hence the second equality; with $\bar{\eta}_S \,{\leq}\, \bar{\eta} \,{\leq}\, \bar{\eta}_O$ and $\bar{r}_O \,{\leq}\, \bar{r}(\bar{\eta}) \,{\leq}\, \bar{r}_S$. Equations \eqref{posVector} and \eqref{displacement} imply that the relevant direction for observation is given by ${-\bf n}$, being the direction of the observer relative to the propagating photon. We define the affine parameter in the background universe, given by $\lambda \,{\equiv}\, {\bf n}\cdot{\bf x}$. Consequently, we have $d\lambda \,{=}\, {-}d\bar{r}$, and therefore,
\begin{equation}\label{eq:dr}
d\bar{\eta} = -d\bar{r} = d\lambda,
\end{equation}
where the first quality holds for a speed of light $c \,{=}\, 1$. Then given \eqref{eq:pertgeod19} and \eqref{eq:dr} we obtain the following 
\begin{align}\label{pert_n0}
\delta{n}^0 \Big\|^S_O &= \Big[\delta{g}_{0\beta}\, \bar{n}^\beta \Big]^S_O + \dfrac{1}{2} \int^{\bar{r}_S}_0{ d\bar{r}\, \delta{g}'_{\alpha\beta}\, \bar{n}^\alpha\bar{n}^\beta } , \\ \label{pert_ni}
\delta{n}^i \Big\|^S_O &= -\Big[\bar{g}^{ij}\delta{g}_{j\beta}\, \bar{n}^\beta \Big]^S_O - \dfrac{1}{2} \bar{g}^{ij} \int^{\bar{r}_S}_0{ d\bar{r}\, \partial_j (\delta{g}_{\alpha\beta})\, \bar{n}^\alpha\bar{n}^\beta } ,
\end{align}
where henceforth, a prime denotes derivative with respect to $\eta$. From \eqref{posVector} we see that $x^\mu(\eta_O) \,{=}\, 0 \,{=}\, \delta{x}^0(\eta_S)$. Thus, the position vector is solely a spatial perturbation: 
\begin{eqnarray}\label{DeviatnVec}
\delta{x}^i(\bar{\eta}_S) &=& \dfrac{1}{2} \int^{\bar{r}_S}_0{ d\bar{r}  (\bar{r}_S-\bar{r}) \left(\bar{g}^{ij}\partial_j\delta{g}_{\alpha\beta} + \delta{g}'_{\alpha\beta}\bar{n}^i \right) \bar{n}^\alpha \bar{n}^\beta } \nonumber\\
&& +\; \int^{\bar{r}_S}_0{ d\bar{r} \left( \bar{g}^{ij}\delta{g}_{j\beta} + \delta{g}_{0\beta}\bar{n}^i\right) \bar{n}^\beta } ,
\end{eqnarray}
where we used \eqref{pert_n0} and \eqref{pert_ni}; $\delta{n}^\mu \,{\equiv}\, \delta{n}^\mu \|^S_O \,{=}\, {-}\delta{n}^\mu \|^O_S$ and, we have integrated by parts once and applied the stationarity condition---which results in terms that vanish or, surface terms which do not contribute to physical solutions (and hence, are discarded).


\section{The Relativistic Galaxy Overdensity}\label{sec:RDP}
Henceforth, a check symbol above a parameter denotes redshift space---being the usual space of cosmological measurements; overbars denote ``background'' quantities: averaged over all solid angles. 

In cosmological surveys, observers usually look at a certain volume $\check{\upsilon}({\bf n},z)$ of the sky through a solid angle $\Omega_{\bf n}$ in a given direction $-{\bf n}$ (see section \ref{sec:TDV}), at a redshift $z$ away. However, apart from the redshift, the solid angle and the intrinsic properties of the observed sources, surveys will generically also depend on cosmic magnification~\cite{Jeong:2011as, Duniya:2015ths, Duniya:2016gcf, Kostelecky:2008iz, Schmidt:2009rh, Schmidt:2010ex, Liu:2013yna, Camera:2013fva, Hildebrandt:2015kcb, Bonvin:2008ni, VanWaerbeke:2009fb, Weinberg:2012es, Duncan:2013haa, Umetsu:2015baa, Gillis:2015caa, Heavens:2011ei} via the apparent flux $\check{F}({\bf n},z)$ (or luminosity) of the observed sources~\citep{Jeong:2011as, Duniya:2015ths, Duniya:2016gcf}. The observed flux of a source is inherently (de)amplified in an inhomogeneous universe. (Hereafter, we use ``magnification'' to mean \emph{magnification of sources} by the amplification of their flux and/or angular size.) Thus, the physical number of galaxies observed in an \emph{arbitrary} survey will depend on two main quantities---which are, the observed sky volume and the apparent flux of the galaxies.

Moreover, a generic sample of cosmic objects in the sky would intrinsically contain both ``magnified'' and ``unmagnified'' fractions, respectively; where the magnified fraction is proportional to the magnification bias. However, the actual measurable quantity, which is the number of objects per unit solid angle per redshift, would consist of all the events together, with no distinctions between these fractions. Nevertheless, these fractions can each be isolated by observations and measured separately, given that the unmagnified fraction is volume-dependent and the magnified fraction is flux-dependent.

\subsection{The Cumulative Number Densities}\label{subsec:CND}
As earlier stated, the physical number $N({\bf n},z)$ of galaxies observed in an \emph{arbitrary} survey will depend on two main quantities which are, the observed volume $\check{\upsilon}({\bf n},z)$ and the apparent flux $\check{F}({\bf n},z)$ of the galaxies---that is, $N({\bf n},z) \,{=}\, N\left(\check{\upsilon}({\bf n},z), \check{F}({\bf n},z)\right)$. 

It should be pointed out that the apparent flux indeed relates directly to the observed sources, and not indirectly through the survey volume. The change $dF$ in apparent flux in a galaxy distribution only occurs from source to source, owing to the difference in flux of the individual sources (of different sizes), and not from any changes in the observed volume or source position. The difference in apparent flux between any two sources is unaffected by the size of the sky volume containing the sources, or by any deformations of the volume. Thus, the apparent flux is independent of the survey volume; with the observed number of sources depending directly on the source apparent flux as on the survey volume.

Therefore, the \emph{ensemble average} number of galaxies per unit volume per unit flux, $\Theta$---which is directly related to the bolometric average number per voxel per flux interval (as measured by observers)---is given by
\begin{eqnarray}\label{Theta}
\Theta = \dfrac{\partial^2 N(\check{\upsilon}, \check{F})}{\partial\check{\upsilon}\partial\check{F}},
\end{eqnarray}
which inherently contains the {\it flux} (or {\it luminosity}) {\it function}. Note that the luminosity function~\cite{Schechter:1976ps, Felten:1977jef, Martinez:2002bk}---which has become ``standard'' and commonplace in cosmological analyses---is only an approximation of $\Theta$: it only accounts for the flux dependence. The luminosity function, often given analytically as an explicit function of luminosity or absolute magnitude, is computed at a fixed volume or distance: in particular, volume and/or galaxy number per unit volume only appears as a constant. Thus, care must be taken to apply the right parameters and properly account for their full nature, in any analysis. Now given \eqref{Theta}, the ensemble average number of galaxies per unit volume, with flux in the infinitesimal range from $\check{F}$ to $\check{F} \,{+}\, d\check{F}$, is given by $\Theta d\check{F}$. Similarly, $\Theta d\check{\upsilon}$ gives the ensemble average number per unit flux, of galaxies at a distance within the volume element from $\check{\upsilon}$ to $\check{\upsilon} \,{+}\, d\check{\upsilon}$. 

The {\it cumulative} number of sources per unit (observed) volume, $\check{n}_{\rm g}$, is 
\begin{eqnarray}\label{def:n_g}
\check{n}_{\rm g}(\check{\upsilon}) = \int^{F_*}_{F_{\rm cut}}{d\check{F}\, \Theta(\check{\upsilon}, \check{F})},
\end{eqnarray}
where $F_{\rm cut} \,{=}\, \bar{F}(m_{\rm cut})$ is the instrument cut-off or threshold flux, with apparent magnitude $m_{\rm cut}$, and $F_* \,{=}\, \bar{F}(m_*)$ is the sample maximum flux, with apparent magnitude $m_*$. (Throughout this work, the apparent magnitude $m$ is taken as being purely background.) Note that the cumulative galaxy number per unit volume is computed at a fixed flux $F_*$, i.e.~$\check{n}_{\rm g}(\check{\upsilon}) \,{=}\, \check{n}_{\rm g}(\check{\upsilon}{\mid}\check{F}{=}F_*)$. In principle, the upper limit in \eqref{def:n_g} may be set at infinity, as is often done in the literature. However, in practice, the upper limit is the maximum flux value in the sample. Sometimes the upper limit is left arbitrarily (e.g. as $F$), but this is only to be understood as denoting a finite maximum value, and not a range of values. (At any finite time, or finite period of time, the observed flux from any source or group of sources must be finite.) Surveys like the SDSS\footnote{\url{http://www.sdss.org/surveys/}} \cite{Eisenstein:2011sa, Albareti:2016xlm} and the LCRS\footnote{\url{http://qold.astro.utoronto.ca/~lin/lcrs.html}} \cite{Shectman:1996km, Lin:1996td}, among others, employ two flux cut-offs, so that only sources with apparent magnitude in a given closed interval are included. 

Similarly, the \emph{cumulative} number of sources per unit (apparent) flux, $\check{n}_F$:
\begin{eqnarray}\label{def:n_F}
\check{n}_F(\check{F}) = \int^{\upsilon_*}_{\upsilon_{\rm min}}{d\check{\upsilon} \,\Theta(\check{\upsilon}, \check{F})},
\end{eqnarray}
where $\upsilon_{\rm min} \,{=}\, \bar{\upsilon}(\bar{r}_{\rm min})$ is the initial volume at a minimum distance $\bar{r}_{\rm min}$ and, $\upsilon_* \,{=}\, \bar{\upsilon}(\bar{r}_*)$ is the final volume at a maximum distance $\bar{r}_*$. Similarly, we see that the cumulative number per unit flux is computed at a fixed volume $\upsilon_*$, i.e.~$\check{n}_F(\check{F}) \,{=}\, \check{n}_F(\check{\upsilon}{=}\upsilon_*{\mid}\check{F})$. Moreover, note that by integrating within fixed limits---in \eqref{def:n_g} and \eqref{def:n_F}---any possible perturbations coming directly from flux change and volume change must vanish. Consequently, any perturbations in $\check{n}_{\rm g}$ and $\check{n}_F$ arise directly via $\Theta$.


\subsection{The Galaxy Number per unit Solid Angle per Redshift}
A volume element vector $\boldsymbol{d\check{\upsilon}}$ and a flux element vector $\boldsymbol{d\check{F}}$ will result in a number of sources, which is prescribed by the resultant number element vector $\boldsymbol{dN}$, given by
\begin{eqnarray}\label{vect-dN1}
\boldsymbol{dN} = \check{n}_{\rm g} \boldsymbol{d\check{\upsilon}} + \check{n}_F \boldsymbol{d\check{F}},
\end{eqnarray}
where $\check{n}_{\rm g}$ and $\check{n}_F$ are scalar functions. We note that, given \eqref{Theta}--\eqref{def:n_F}, the vector component $\check{n}_{\rm g} \boldsymbol{d\check{\upsilon}}$ is independent of flux and, the vector component $\check{n}_F \boldsymbol{d\check{F}}$ is independent of volume. Thus, the magnitude $dN$ of the resultant number (element) vector~\eqref{vect-dN1}, is given by
\begin{eqnarray}\label{vect-dN2}
dN^2 &=& \check{n}^2_{\rm g}\, (\boldsymbol{d\check{\upsilon} \cdot d\check{\upsilon}}) + \check{n}^2_F\, (\boldsymbol{d\check{F} \cdot d\check{F}}) ,\\
&=& \left\lbrace\Big(\check{n}_{\rm g} \check{\cal V}\Big)^2 + \Big(\check{n}_F \check{\cal F} \check{q}_{\cal X}\Big)^2 \right\rbrace \Big[d\Omega_{\bf n} dz\Big]^2,\nonumber\\ \label{vect-dN3}
&{\equiv}& \check{\cal N}_{\rm g}^2\; \Big[d\Omega_{\bf n} dz\Big]^2,
\end{eqnarray} 
where $dN^2 \,{=}\, \boldsymbol{dN \cdot dN}$ and, we have
\begin{eqnarray}\label{vect-dV}
\boldsymbol{d\check{\upsilon}} = \boldsymbol{e}_\upsilon d\check{\upsilon} = \boldsymbol{e}_\upsilon \check{\cal V} d\Omega_{\bf n} dz,
\end{eqnarray}
with $\boldsymbol{e}_\upsilon$ being the (orthonormal) volume unit vector, $d\check{\upsilon}$ being the magnitude of $\boldsymbol{d\check{\upsilon}}$ and, $\check{\cal V} \,{=}\, \partial\check{\upsilon} / (\partial{\Omega_{\bf n}} \partial{z})$ being the volume per unit solid angle per redshift. We assume a general scenario where the survey also pertains the apparent size $\check{\cal X}$ of the sources, hence the observed flux depends on the apparent size of the sources, i.e. $\check{F}({\bf n},z) \,{=}\, \check{F}(\check{\cal X}({\bf n},z))$, and we have 
\begin{eqnarray}\label{vect-dF}
\boldsymbol{d\check{F}} = -\boldsymbol{e}_F d\check{F} = -\boldsymbol{e}_F \check{\cal F} \check{q}_{\cal X} d\Omega_{\bf n} dz,
\end{eqnarray}
where $\boldsymbol{e}_F$ is the (orthonormal) flux unit vector, $d\check{F}$ is the magnitude of $\boldsymbol{d\check{F}}$, $\check{\cal F} \,{=}\, \partial\check{F}/\partial\check{\cal X}$ is the flux change per source size interval $d\check{\cal X}$ (i.e.~flux per unit size) and, $\check{q}_{\cal X} \,{=}\, \partial\check{\cal X}/(\partial{\Omega_{\bf n}} \partial{z})$ is the source size per unit solid angle per redshift. (Notice that the product $\check{\cal F}\check{q}_{\cal X}$, gives the flux per unit solid angle per redshift.) The quantity $\check{n}_{\rm g} d\check{\upsilon}$ (being the magnitude of the vector $\check{n}_{\rm g} \boldsymbol{d\check{\upsilon}}$) is actually the number of galaxies contained in the physical volume element $d\check{\upsilon}$---arising solely owing to the volume variation---and, $\check{n}_F d\check{F}$ (being the magnitude of the vector $\check{n}_F \boldsymbol{d\check{F}}$) is the number of galaxies arising solely owing to the change $d\check{F}$ in apparent flux.

Equation \eqref{vect-dN1} provides a correct and consistent approach for incorporating the magnification bias in the relativistic overdensity of a generic galaxy survey: the cosmic magnification---apart from the redshift, the solid angle, and the intrinsic properties of the sources---surfaces via the flux (or luminosity) of the observed sources. Until this work, all previous works (including \cite{Bonvin:2011bg, Challinor:2011bk, Jeong:2011as, Duniya:2015ths, LopezHonorez:2011cy, Bonvin:2014owa, Alonso:2015uua, Renk:2016olm, Durrer:2016jzq}) on the relativistic galaxy overdensity have failed to properly take into account the ensemble average $\Theta$ and/or the fact that it is both volume and flux (or luminosity) dependent.

Thus, given \eqref{vect-dN1}--\eqref{vect-dF} we have the (physical) number of galaxies per unit solid angle per redshift, $\check{\cal N}_{\rm g}$---what observers actually measure---given by \eqref{tildeNg}:
\begin{eqnarray}\label{tildeNg}
\check{\cal N}_{\rm g}({\bf n},z) &=& \left\{\Big[\check{n}_{\rm g}({\bf n},z) \check{\cal V}({\bf n},z)\Big]^2 + \Big[\check{n}_F({\bf n},z) \check{\cal F}({\bf n},z) \check{q}_{\cal X}({\bf n},z)\Big]^2 \right\}^{\frac{1}{2}},\nonumber\\
&=& \; \check{\left< {\cal N}_{\rm g} \right>} (\bar{z}) \left\{ 1 + \epsilon^2_{\cal V}(\bar{z})\left[ \dfrac{\delta\check{n}_{\rm g}({\bf n},z)}{\bar{n}_{\rm g}(\bar{z})} + \dfrac{\delta\check{\cal V}({\bf n},z)}{\bar{\cal V}(\bar{z})}\right] \right. \nonumber\\
&&\hspace{2cm} + \left. \epsilon^2_{\cal F}(\bar{z}) {\cal Q}(\bar{z}) \dfrac{\delta\check{\cal F}({\bf n},z)}{\bar{\cal F}(\bar{z})} \right\},
\end{eqnarray}
where the redshift distribution of the galaxies is obtained by averaging over all solid angles: $d\bar{N}(\bar{z}) \,{=}\, \check{\left< {\cal N}_{\rm g} \right>} (\bar{z}) dz$; 
\begin{eqnarray}\label{calNsqr}
\check{\left< {\cal N}_{\rm g} \right>}^2 = \left(\bar{n}_{\rm g}\bar{\cal V}\right)^2 \,{+}\, \left(\bar{n}_F \bar{\cal F} \bar{q}_{\cal X}\right)^2,
\end{eqnarray}
with $\check{\left< {\cal N}_{\rm g} \right>}$ being the average galaxy distribution density in redshift space. Thus \eqref{calNsqr} gives the identity
\begin{eqnarray}\label{fracSum}
\epsilon^2_{\cal V}(\bar{z}) + \epsilon^2_{\cal F}(\bar{z}) = 1,
\end{eqnarray}
where $\epsilon_{\cal V} \,{\equiv}\, \bar{n}_{\rm g}\bar{\cal V} / \check{\left< {\cal N}_{\rm g} \right>}$ and $\epsilon_{\cal F} \,{\equiv}\, \bar{n}_F \bar{\cal F} \bar{q}_{\cal X} / \check{\left< {\cal N}_{\rm g} \right>}$ are the fractional contributions by the change in survey volume and the change in apparent flux, respectively, in the average galaxy distribution density. That is, if the survey volume and the apparent flux are variable, then the change in volume will contribute a certain fraction $\epsilon_{\cal V}$ in the average distribution density and, similarly, the variation in flux will also contribute a fraction $\epsilon_{\cal F}$. 

The parameter ${\cal Q}$ in \eqref{tildeNg} is the cosmic \emph{magnification bias} \citep{Jeong:2011as, Duniya:2015ths, Duniya:2016gcf, Kostelecky:2008iz, Schmidt:2009rh, Schmidt:2010ex, Liu:2013yna, Camera:2013fva, Montanari:2015rga, Hildebrandt:2015kcb, Bartelmann:1995vxp, Ziour:2008awn, MoradinezhadDizgah:2016pqy, Vallinotto:2007mf, Narayan:1989apj}, given by
\begin{eqnarray}\label{b_M}
{\cal Q} \equiv \left. \dfrac{\partial\ln\bar{\cal N}_F}{\partial\ln\bar{\cal F}} \right\|_{\bar{z}} = \dfrac{1}{2} \left(2 + 2s_{\cal X} - 5s_F\right),
\end{eqnarray}
where $\bar{\cal N}_F \,{\equiv}\, \bar{n}_F \bar{\cal F} \bar{q}_{\cal X}$ and, $s_{\cal X} \,{\equiv}\, \partial\ln\bar{q}_{\cal X}/\partial\ln\bar{\cal F}$ is the \emph{size bias} \citep{Schmidt:2009rh, Schmidt:2010ex, Liu:2013yna}; $s_F \,{\equiv}\, {-}(2/5)\partial\ln\bar{n}_F/\partial\ln\bar{\cal F} \,{=}\, \partial\log_{10}\bar{n}_F/\partial{m}$ is the \emph{flux bias} \citep{Lin:1996td}, and $m$ is the apparent magnitude. 

For a given source, the surface brightness ${\cal S} \,{\propto}\, (\check{F} / \check{\cal X}^2)$ is a constant and, is unaffected by cosmic amplification (or magnification) of the source~\citep{Schmidt:2009rh}. This implies that $\bar{F} \propto \bar{\cal X}^2$, and we get $s_{\cal X} \,{=}\, \partial \ln\bar{q}_{\cal X} / \partial \ln\bar{\cal X}$. However, if the survey does not pertain the apparent size of the source, we will have in \eqref{tildeNg} $\check{n}_F\check{\cal F}\check{q}_{\cal X}\, {\to}\, \check{n}_F\check{\cal F}$; thus, $\bar{\cal N}_F \,{=}\, \bar{n}_F \bar{\cal F}$, where $\check{\cal F}$ becomes the flux per unit solid angle per redshift. Consequently, $s_{\cal X} \,{=}\, 0$ and we have $2{\cal Q} \,{=}\, 2 \,{-}\, 5s_F$. 

Equation \eqref{b_M} therefore generalizes the cosmic magnification bias---in the light of previous works.


\subsection{The Observed Galaxy Overdensity}\label{sec:ObsDelta_g}
The total observed, relativistic galaxy overdensity is 
\begin{eqnarray}\label{ObzDelta} 
\Delta^{\rm obs}_{\rm g}({\bf n},z) &\equiv & \dfrac{\check{\cal N}_{\rm g}({\bf n},z) - \check{\left< {\cal N}_{\rm g} \right>}(\bar{z})}{ \check{\left< {\cal N}_{\rm g} \right>}(\bar{z})},\\ \label{ObzDelta2} 
&=& \epsilon^2_{\cal V}(\bar{z}) \Big(\check{\delta}_{\rm g}({\bf n},z) + \check{\delta}_{\cal V}({\bf n},z)\Big) 
+ \epsilon^2_{\cal F}(\bar{z})\, {\cal Q}(\bar{z})\, \check{\delta}_{\cal M}({\bf n},z),
\end{eqnarray} 
where we used \eqref{tildeNg}, with $\check{\delta}_{\rm g} \,{\equiv}\, \delta\check{n}_{\rm g} /\bar{n}_{\rm g}$ being the number density contrast and, $\check{\delta}_{\cal V} \,{\equiv}\, \delta{\check{\cal V}} / \bar{\cal V}$ is the volume density contrast. In the last line in \eqref{ObzDelta2} we have used that for magnified sources, the flux per unit source size---in the background---is proportional to the underlying magnification per unit source size, i.e.~$\bar{\cal F} \,{\propto}\, \bar{\cal M}$ where $\mathcal{\tilde M}$ is the magnification per unit source size, and we have $\partial\ln\bar{\cal F} \,{=}\, \partial\ln\bar{\cal M}$. Consequently, we have the magnification density contrast, given by $\check{\delta}_{\cal M} \,{\equiv}\, \delta{\check{\cal M}} / \bar{\cal M} \,{=}\, \delta{\check{\cal F}} / \bar{\cal F}$. 
 
Equation \eqref{ObzDelta2} establishes that although the number density contrast $\check{\delta}_{\rm g}$ and the volume density contrast $\check{\delta}_{\cal V}$ add up directly---without any cofactors---in the observed galaxy overdensity, it does not follow in the same manner when it comes to including the magnification density contrast $\check{\delta}_{\cal M}$. This is tricky. One may be deceived by the fact that $\check{\delta}_{\rm g}$ and $\check{\delta}_{\cal V}$ (being physical quantities) sum directly and thus, think that the magnification term should also be added directly, i.e.~$\Delta^{\rm obs}_{\rm g} \,{=}\, \check{\delta}_{\rm g} \,{+}\, \check{\delta}_{\cal V} \,{+}\, {\cal Q}\check{\delta}_{\cal M}$, as is the case in previous works (e.g.~\cite{Yoo:2009au, Yoo:2010ni, Challinor:2011bk, Jeong:2011as, LopezHonorez:2011cy, Duniya:2015ths, Alonso:2015uua, Renk:2016olm}). However, doing so is incorrect and inconsistent: it does not properly take into account, the nature of the ensemble average (galaxy) number per unit volume per unit flux, $\Theta$, which has partial dependence on both volume and flux. As already discussed under \eqref{Theta}, the so-called luminosity function, which has become commonplace in cosmological analyses, is merely an approximation of $\Theta$: it only accounts for the flux dependence.

\subsubsection{The density distortion}\label{subsec:RSD} 
To compute $\check{\delta}_{\rm g}({\bf n},z)$, which is in redshift space, we first relate it to the real-space distortion $\delta_{\rm g}({\bf x},\eta)$---by taking a gauge transformation from real space to redshift space: 
\begin{equation}\label{delta_g2}
\check{\delta}_{\rm g}({\bf n},z) = \delta_{\rm g}({\bf n},z) - \dfrac{d\ln\bar{n}_{\rm g}}{d\bar{z}}\,\delta{z}({\bf n},z),
\end{equation} 
where the first term on the right hand side of \eqref{delta_g2} denotes the real-space number distortion expressed as a function of redshift, and we used $\delta{\eta} \,{=}\, (\partial\bar{\eta}/\partial\bar{z})\delta{z}$. Then we have,
\begin{eqnarray}\label{delta:g3}
\dfrac{d\ln\bar{n}_{\rm g}}{d\ln(1+\bar{z})} = 3 - b_e, 
\end{eqnarray}
where $a \,{=}\, 1/(1+\bar{z})$ and, $b_e \,{=}\, b_e(a)$ is the galaxy \emph{evolution bias} \citep{Jeong:2011as, Duniya:2015ths}, given by
\begin{equation}\label{b_e}
b_e \;\equiv\; \left. \dfrac{\partial\ln(a^3\bar{n}_{\rm g})}{\partial\ln(a)} \right\|_a, 
\end{equation}
with $a^3 \bar{n}_{\rm g}$ being the background comoving galaxy number density. Thus, the density distortion \eqref{delta_g2} becomes
\begin{equation}\label{delta-g3}
\check{\delta}_{\rm g} ({\bf n},z) = \delta_{\rm g}({\bf n},z) + \Big[b_e(\bar{z}) - 3\Big]\,\dfrac{\delta{z}}{1+\bar{z}}. 
\end{equation}

A photon moving along ${\bf n}$ from $S$ will be seen by $O$, with fundamental 4-velocity $\tilde{u}^\mu$ being govern by the metric \eqref{conformal-ds2}, to be approaching with an energy $E \,{=}\, {-}\tilde{n}^{\mu} \tilde{u}_{\mu}$. (See discussion under \eqref{null}, noting that $\tilde{u}_\mu$ is a conformal Killing vector.) In a perturbed universe, the 4-velocity of a particle is perturbed by $\tilde{u}^\mu \,{=}\, \bar{u}^{\mu} + \delta\tilde{u}^{\mu}$, which obeys the normalization given by $\tilde{u}^\mu \tilde{u}_\mu \,{=}\, {-}1$. Thus, 
\begin{eqnarray}\label{vels}
\tilde{u}^\mu &=& a^{-1}u^\mu = a^{-1}\left(1 -\phi,~v^{\mid  i}\right),\nonumber\\ 
\tilde{u}_\mu &=& a u_\mu = a\left(-1 -\phi, v_{\mid  i}+B_{\mid  i}\right),
\end{eqnarray}
where $v^i \,{=}\, v^{\mid  i}$ is the `peculiar' velocity, and $v$ is the velocity potential (a scalar). The peculiar velocity is irrotational, i.e.~the curl of the velocity $\boldsymbol{\nabla} \,{\wedge}\, {\bf v} \,{=}\, 0$, hence the velocity can be given as the gradient of $v$. 

The redshift $z$ in a photon---being the fractional change in energy of the photon in propagating through the gravitational field from $S$ to $O$---is given by
\begin{equation}\label{redshift}
1+z = \dfrac{\left(\tilde{n}^{\mu}\tilde{u}_{\mu}\right)_S}{\left(\tilde{n}^{\mu}\tilde{u}_{\mu}\right)_O}  = \dfrac{a_O}{a_S} \left[1 + \delta\left({n}^\mu {u}_\mu\right) \Big\|^O_S \right],
\end{equation}
where we used \eqref{n:Conf} and \eqref{vels}. By using $z \,{=}\, \bar{z}+\delta{z}$, we have the redshift perturbation from \eqref{redshift}, given by  
\begin{equation}\label{pertRedshift}
\dfrac{\delta{z}}{1+\bar{z}} = \delta\left({n}^\mu {u}_\mu\right) \Big\|^O_S = \Big[\bar{n}^iv_{\mid  i} +\bar{n}^iB_{\mid  i} -\delta{n}^0 -\phi \Big]^O_S .
\end{equation}
Hereafter, we will work with the gauge-invariant potentials $\Phi$, $\Psi$ and $V$ (see Appendix \ref{sec:metric}), given by
\begin{eqnarray}\label{Phi}
\Phi &=& \phi - {\cal H}\sigma - {\sigma}',\\ \label{Psi}
\Psi &=& \psi + {\cal H}\sigma, \\ \label{Vel}
V &=& v + E' ,
\end{eqnarray}
where $\sigma \,{=}\, {-}B + E'$. By using \eqref{pert_n0} and \eqref{pertRedshift}--\eqref{Vel} (see Appendix~\ref{appendix:DMD} for details), the redshift \eqref{redshift}, is given by
\begin{eqnarray}\label{eq:33}
1+z &=& (1+\bar{z})\left[1 + \left(\Phi + \Psi + {\bf n}\cdot{\bf V} -\psi\right) \Big\|^O_S + \int^0_{\bar{r}_S}{d\bar{r} \Big(\Phi' + \Psi'\Big) } \right], 
\end{eqnarray} 
where ${\bf V} \,{=}\, \partial_i V$ is the gauge-invariant velocity, with $V$ being the associated velocity potential---given by \eqref{Vel}.

Thus, the redshift-space density distortion \eqref{delta:g3}, is
\begin{equation}\label{delta:g4}
\check{\delta}_{\rm g} = \Delta_\psi + 3\Psi + (3-b_e)\left[ \Phi + {\bf n}\, {\cdot} {\bf V} + \int^{\bar{r}_S}_0{d\bar{r} \left(\Phi' + \Psi'\right)}\right] , 
\end{equation} 
where (henceforth) we drop the limit from all parameters outside integrals: hence they denote the relative value between the source and the observer positions, respectively. (Note that $b_e$ appearing in \eqref{delta:g3}--\eqref{delta-g3} and \eqref{delta:g4}, is incorrectly omitted in \cite{Bonvin:2011bg, Durrer:2016jzq}.) The parameter $\Delta_\psi \,{\equiv}\, \delta_{\rm g} \,{-}\, 3\psi \,{-}\, b_e {\cal H}\sigma$ denotes the density perturbation in uniform curvature gauge, and it evaluates to
\begin{equation}\label{flatCurvDelta}
\Delta_\psi = \Delta_{\rm g} +\left(3-b_e\right){\cal H}V -3\Psi,
\end{equation}
where $\Delta_{\rm g} \,{\equiv}\, \delta_{\rm g} \,{+}\, (v_{\rm g} \,{+}\, B) \bar{n}_{\rm g}'/\bar{n}_{\rm g}$ is the comoving galaxy number overdensity---which is gauge-invariant, and $\bar{n}'_{\rm g} \,{=}\, {-}{\cal H} \left(3-b_e\right) \bar{n}_{\rm g}$, with $b_e$ being given by \eqref{b_e}; $\Psi$ and $V$ are as given by \eqref{Psi} and \eqref{Vel}, respectively, and ${\cal H} \,{=}\, a'/a$ is the (conformal) Hubble parameter.

\subsubsection{The volume distortion}\label{subsec:VolDistortn}%
In a flat background FRW universe, geodesics reduce to straight lines. Hence the photon emission angles, $\theta_S$ and $\vartheta_S$, are equal to the corresponding observed angles, $\theta_O$ and $\vartheta_O$, where $\theta$ and $\vartheta$ are the zenith and the azimuthal angles, respectively. However, in a perturbed universe these angles become perturbed, given by
\begin{equation}\label{pertAngles}
\theta_S = \theta_O + \delta{\theta}, \quad \vartheta_S = \vartheta_O + \delta{\vartheta},
\end{equation}
where we use $\delta{\theta}_O \,{=}\, 0 \,{=}\, \delta{\vartheta}_O$. The perturbations in the redshift \eqref{eq:33} and in the angles \eqref{pertRedshift} result in the survey volume becoming perturbed: leading to a distortion $\tilde{\delta}_{_{\cal V}}$. 

By taking a gauge transformation from real space to redshift space, we have
\begin{equation}\label{delta:v2}
\check{\delta}_{\cal V}({\bf n},z) = \delta_{\cal V}({\bf n},z) - \dfrac{d\ln\bar{\cal V}}{d\bar{z}}\,\delta{z}({\bf n},z),
\end{equation}
where $\delta_{\cal V}$ denotes the volume distortion in real space ${\bf x}$. An infinitesimal real volume element containing a source with 4-velocity $\tilde{u}^\mu$, given by \eqref{vels}, is 
\begin{eqnarray}\label{eq:34}
d{\bf \upsilon} &=& \sqrt{-\tilde{g}}\; \epsilon_{\mu\nu\alpha\beta} \tilde{u}^\mu d\tilde{x}^{\nu}d\tilde{x}^{\alpha}d\tilde{x}^{\beta}, \\ \label{eq:34b}
&\equiv & {\cal V}\left(z,\theta_O ,\vartheta_O\right) dz\, d\theta_O d\vartheta_O ,
\end{eqnarray} 
where $\epsilon_{\mu\nu\alpha\beta}$ is the alternating tensor, and
\begin{equation}\label{eq:35}
{\cal V} \,{=}\, \sqrt{-\tilde{g}}\; \epsilon_{\mu\nu\alpha\beta}\, \tilde{u}^\mu \dfrac{\partial \tilde{x}^\nu}{\partial z} \dfrac{\partial \tilde{x}^\alpha}{\partial\theta_S} \dfrac{\partial \tilde{x}^\beta}{\partial \vartheta_S} \left. \left\| \dfrac{\partial (\theta_S,\vartheta_S)}{\partial \left(\theta_O ,\vartheta_O\right)} \right. \right\| ,
\end{equation}
with ${\cal V}$ being the real-space volume density. Thus, after some calculations (see Appendix~\ref{appendix:Volpert}), we have
\begin{equation}\label{eq:41}
{\cal V} = a^3 (1+\phi-3D) \dfrac{dr}{dz}r^2 \| J\| \sin{\theta}_S - a^3 \left(\phi\dfrac{d\bar{r}}{d\bar{z}} + v_r\dfrac{d\bar{\eta}}{d\bar{z}} \right)\bar{r}^2 \sin{\theta}_O ,
\end{equation}
where $J \equiv \partial (\theta_S, \vartheta_S) / \partial\left(\theta_O,\vartheta_O\right)$, with $r \,{=}\, \bar{r} \,{+}\,\delta{r}$ and, 
\begin{eqnarray}\label{eq:39}
\sqrt{-\tilde{g}} = a^4\left(1 + \dfrac{1}{2}\delta{g}^\alpha\/_\alpha\right) = a^4 (1+\phi - 3D) ,
\end{eqnarray}
where $\tilde{g}$ is the determinant of $\tilde{g}_{\mu\nu}$, which has components given by the metric \eqref{conformal-ds2}. Then by applying the relevant equations and simplifying, we have
\begin{equation}\label{eq:42}
\dfrac{{\cal V}}{\bar{\cal V}} =  1 - 3D + v_r  + 2\dfrac{\delta{r}}{\bar{r}} - \dfrac{d}{d\lambda}\delta{r}  + \dfrac{a}{{\cal H}} \dfrac{d}{d\lambda}\delta{z} + \left(\cot{\theta_O} +\partial_{\theta}\right)\delta{\theta} + \partial_{\vartheta}\delta{\vartheta}  ,
\end{equation}
where $\bar{\cal V}(\bar{z}) \,{\equiv}\, a^4\bar{r}(\bar{z})^2{\cal H}(\bar{z})^{-1}\sin{\theta}_O$. We used  \eqref{eq:dr} and, $d\bar{r} /d\bar{z} \,{=}\, {-}d\bar{\eta} / d\bar{z} \,{=}\, a / {\cal H}$: to lowest order along the photon geodesic. (Note that $v_r \,{=}\, e^i_r v_i$, where $e^i_r$ is the radial unit vector in polar coordinates.) We assumed $\|\delta\theta\| \,{\ll}\, 1$ and, given \eqref{pertAngles}, we have $\sin{\theta_S} \,{=}\, (1 + \delta{\theta}\cot{\theta_O}) \sin{\theta_O}$; with 
\begin{equation}\label{eq:37}
\left. \left\| J \right. \right\| = 1 +\partial_{\theta}\delta{\theta} + \partial_{\vartheta}\delta{\vartheta} ,
\end{equation}
where $\partial_{\theta} \,{=}\, \partial/\partial{\theta}$ (similarly for $\vartheta$). 

By using the fact that the real-space volume distortion in \eqref{delta:v2}, is given by 
\begin{equation}\label{delta_V2}
\delta_{\cal V}({\bf n},z) = \dfrac{{\cal V}({\bf n},z) -\bar{\cal V}(\bar{z})}{\bar{\cal V}(\bar{z})},
\end{equation}
and by employing \eqref{delta:v2} and \eqref{eq:42}, we have 
\begin{eqnarray}\label{delta_V3}
\check{\delta}_{\cal V} &=&  -3D + v_r + 2\dfrac{\delta{r}}{\bar{r}} - \dfrac{d\delta{r}}{d\lambda} + \dfrac{a}{{\cal H}} \dfrac{d\delta{z}}{d\lambda} + \partial_{\vartheta}\delta{\vartheta}  \nonumber\\
&& +\; \left(\cot{\theta_O} + \partial_{\theta}\right)\delta{\theta} + a\left(4 -\dfrac{2}{\bar{r} {\cal H}} -\dfrac{{\cal H}'}{{\cal H}^2} \right) \delta{z},
\end{eqnarray}
where, 
\begin{equation}
\dfrac{d\ln\bar{\cal V}}{d\ln(1+\bar{z})} = -4  + \dfrac{2}{\bar{r} {\cal H}} + \dfrac{{\cal H}'}{{\cal H}^2} .
\end{equation}
In order to get the physical expression for the volume distortion \eqref{delta_V3}, we need to compute $\delta{r}$, $\delta{\theta}$ and $\delta{\vartheta}$. 

Consider a given hypothetical sphere in spacetime, in which the observer is at the centre and the source is on the circumference. Then the deviation 4-vector in the polar coordinates ${x}^\mu_p$ may be related (to first order) to that of a Cartesian coordinates $x^\mu$, by 
\begin{equation}\label{polCartTrans}
\delta{x}^\mu_p = \dfrac{\partial{x}^\mu_p}{\partial{x}^\nu} \delta{x}^\nu = \delta^\mu\/_\nu\, \delta{x}^\nu ,
\end{equation}
where only the spatial component of the position vectors are relevant to our calculations. The spatial displacement vector of a particle in polar coordinates, is given by
\begin{equation}\label{polarDevVect} 
\boldsymbol{\delta{x}}_p = \delta{r}\, {\pmb e}_r + \bar{r}\, \delta{\theta}\, {\pmb e}_{\theta} + \bar{r}\sin{\theta} \, \delta{\vartheta}\, {\pmb e}_{\vartheta}, 
\end{equation}
where ${\pmb e}_r$, ${\pmb e}_{\theta}$ and ${\pmb e}_{\vartheta}$ are the orthonormal unit vectors of the polar coordinates; with ${\pmb e}_{\theta} \,{=}\, \partial_{\theta} {\pmb e}_r$ and $ {\pmb e}_{\vartheta} \sin{\theta} \,{=}\, \partial_{\vartheta} {\pmb e}_r$. 

Given \eqref{polarDevVect}, we have $\delta{r}$, $\delta{\theta}$ and $\delta{\vartheta}$, given by
\begin{align}\label{VectCompts} 
\delta{r} = {\pmb e}_{r}\cdot\boldsymbol{\delta{x}_p},~~\bar{r}\,\delta{\theta} = {\pmb e}_{\theta}\cdot \boldsymbol{\delta{x}_p},~~\bar{r}\sin{\theta}\,\delta{\vartheta} = {\pmb e}_{\vartheta}\cdot \boldsymbol{\delta{x}_p} ,
\end{align}
where a dot denotes the standard scalar product. Moreover, the components of the Laplacian in polar coordinates are given by
\begin{align}\label{LapsCompts} 
\partial_r = -\bar{n}^i \partial_i,\quad \dfrac{1}{\bar{r}}\partial_{\theta} = e^i_{\theta} \partial_i,\quad \dfrac{1}{\bar{r}\sin{\theta}} \partial_{\vartheta} = e^i_{\vartheta}\partial_i ,
\end{align}
where henceforth, ${e}^i_r \,{=}\, {-}\bar{n}^i$: resulting in $v_r \,{=}\,  {-}{\bf n} \cdot {\bf v}$. 

Therefore, given \eqref{DeviatnVec}, \eqref{VectCompts} and \eqref{LapsCompts}, we have
\begin{equation}\label{delta_r2}
\delta{r} = -\bar{n}_i\, \delta{x}^i = -\dfrac{1}{2} \int^{\bar{r}_S}_{0}{ d\bar{r}\, \delta{g}_{\alpha\beta} \, \bar{n}^\alpha \bar{n}^\beta },
\end{equation}
where we have integrated by parts once and set surface terms to vanish. Similarly, given \eqref{DeviatnVec}, \eqref{VectCompts} and \eqref{LapsCompts}: 
\begin{eqnarray}\label{delta_theta}
\bar{r}_S\,\delta{\theta} = \int^{\bar{r}_S}_0{ d\bar{r}\,\left[\delta{g}_{j\beta} \, e^j_{\theta}\, \bar{n}^\beta + \dfrac{1}{2} (\bar{r}_S - \bar{r})\, e^j_{\theta}\partial_j \left(\delta{g}_{\alpha\beta} \right)\bar{n}^\alpha \bar{n}^\beta \right]} ,
\end{eqnarray}
and,
\begin{eqnarray}\label{delta_vartheta}
\bar{r}_S\,\sin{\theta_O}\,\delta{\vartheta} = \int^{\bar{r}_S}_0{ d\bar{r}\,\left[ \delta{g}_{j\beta} \, e^j_{\vartheta}\, \bar{n}^\beta  + \dfrac{1}{2} (\bar{r}_S - \bar{r})\, e^j_{\vartheta}\partial_j \left(\delta{g}_{\alpha\beta} \right) \bar{n}^\alpha \bar{n}^\beta \right] }.
\end{eqnarray}
By using \eqref{eq:dr} and \eqref{delta_r2}, and applying the appropriate substitutions, we have 
\begin{eqnarray}\label{ddr}
\dfrac{d\delta{r}}{d\lambda} &=& \dfrac{1}{2}\delta{g}_{\alpha\beta}\, \bar{n}^\alpha \bar{n}^\beta =  -\left(\Phi + \Psi\right) + \dfrac{dB}{d\lambda} + \dfrac{d^2E}{d\lambda^2} -2\dfrac{dE'}{d\lambda},
\end{eqnarray}
where we used \eqref{Phi} and \eqref{Psi} and, for a given scalar $X$, we have $dX/d\lambda = X' + \bar{n}^i\partial_i{X}$. After some lengthy, but straightforward calculations (see Appendix~\ref{appendix:DMD}), we have 
\begin{align}\label{dVol}
\left(\cot{\theta_O} + \partial_{\theta}\right) \delta{\theta} + \partial_{\vartheta}\delta{\vartheta} = - \Big[\nabla^2_\perp{E}\Big]^S_O + \int^{\bar{r}_S}_0{ d\bar{r} \left(\bar{r} \,{-}\, \bar{r}_S\right) \dfrac{\bar{r}}{\bar{r}_S} \nabla^2_\perp \left(\Phi + \Psi\right)} ,
\end{align}
where $\nabla^2_{\perp} \,{\equiv}\, \nabla^2 \,{-}\, \partial^2_r \,{-}\, 2\bar{r}^{-1} \partial_r$ is the Laplacian in the source plane, i.e. the plane transverse to the line of sight.

By applying $\partial_r \,{=}\, {-}\bar{n}^i \partial_i$, which follows from our choice of $\boldsymbol{e}_r$ [see \eqref{LapsCompts}]: 
\begin{align}\label{nabPerp}
\nabla^2_{\perp}E = \nabla^{2}E - \dfrac{d^2E}{d\lambda^2} + 2\dfrac{dE'}{d\lambda} - E'' + \dfrac{2}{\bar{r}} \left[\dfrac{dE}{d\lambda} -E'\right] .
\end{align}
Thus, after applying the relevant equations, we have the physical expression of the volume distortion, given by
\begin{eqnarray}\label{obs-deltaV}
\check{\delta}_{\cal V} &=& \dfrac{1}{{\cal H}} \partial_r\left({\bf n}\,{\cdot} {\bf V}\right) + \int^{\bar{r}_S}_0{d\bar{r} \left(\bar{r} - \bar{r}_S\right)\dfrac{\bar{r}}{\bar{r}_S} \nabla^2_\perp \left(\Phi + \Psi\right)} \nonumber\\
 &&+\; \left(\dfrac{2}{\bar{r}_S {\cal H}} + \dfrac{{\cal H}'}{{\cal H}^2}\right) \left[ \Phi + {\bf n}\,{\cdot} {\bf V} + \int^{\bar{r}_S}_0{d\bar{r} \left(\Phi' + \Psi' \right) } \right] \nonumber\\
&&+\; \dfrac{2}{\bar{r}_S}\int^{\bar{r}_S}_0{ d\bar{r} \left(\Phi + \Psi\right) } - 3\int^{\bar{r}_S}_0{ d\bar{r} \left(\Phi' + \Psi'\right) } \nonumber\\
&&+\; \dfrac{1}{{\cal H}}\Psi' - 2\left(\Phi + \Psi\right) - 3\, {\bf n}\,{\cdot} {\bf V} ,
\end{eqnarray}
where we used the Euler equation for the conservation of momentum (density), given by
\begin{equation}\label{Euler-eqn}
{\bf n}\,{\cdot} {\bf V}' + {\cal H}\, {\bf n}\,{\cdot} {\bf V} = \partial_r \Phi,
\end{equation}
with galaxies (hence, pressureless matter) taken to move along geodesics. The Euler equation~\eqref{Euler-eqn} is for non-interacting dark sector; for an interacting dark sector, the Euler equation may get modified (see e.g.~\cite{Duniya:2015nva}). Equation \eqref{obs-deltaV} gives the distortion that arises in the observed volume of galaxy redshift surveys.

Given \eqref{delta:g4}, \eqref{flatCurvDelta} and \eqref{obs-deltaV}, the observed (relativistic) \emph{number-count overdensity}, is given 
\begin{eqnarray}\label{def:Delta_n}
\Delta^{\rm obs}_n({\bf n},z) &\;\equiv\;& \check{\delta}_{\rm g}({\bf n},z) + \check{\delta}_{\cal V}({\bf n},z), \\ \label{Delta_n}
&=& \Delta_{\rm g}({\bf n},z) + \dfrac{1}{{\cal H}} \partial_r \left( {\bf n}\,{\cdot} {\bf V}\right) ({\bf n},z) \nonumber\\
&&+\; \int^{\bar{r}_S}_0{d\bar{r} \left(\bar{r} - \bar{r}_S\right)\dfrac{\bar{r}}{\bar{r}_S} \nabla^2_\perp \Big(\Phi + \Psi\Big)({\bf n},\bar{r})}\nonumber\\
&&+\;  \left(3 - b_e\right){\cal H}V({\bf n},z) + \Phi({\bf n},z) - 2\Psi({\bf n},z) \nonumber\\
&&+\; \dfrac{1}{{\cal H}}\Psi'({\bf n},z) + \dfrac{2}{\bar{r}_S}\int^{\bar{r}_S}_0{d\bar{r} \Big(\Phi + \Psi\Big)({\bf n},\bar{r})} \nonumber\\
&&+ \left(\dfrac{{\cal H}'}{{\cal H}^2} + \dfrac{2}{\bar{r}_S {\cal H}} - b_e\right) \Big[ \Phi({\bf n},z) + {\bf n}\,{\cdot} {\bf V}({\bf n},z) \nonumber\\
&& \hspace{1.2cm} + \int^{\bar{r}_S}_0{d\bar{r} \Big(\Phi' + \Psi' \Big)({\bf n},\bar{r}) } \Big] .
\end{eqnarray}
The first line in \eqref{Delta_n} contains the galaxy comoving number overdensity and, the Kaiser redshift-space distortion term. The second line gives the gravitational lensing term. The first two lines together constitute the dominant terms in the observed number-count overdensity. Whereas, the time-delay term (integral term in the fourth line), the integrated Sachs-Wolfe (ISW) term (integral term in square brackets), the Doppler term (velocity term in square brackets) and, the potential (difference) term (rest of the terms), together constitute the so-called `relativistic corrections' in the number-count overdensity.

\subsubsection{The magnification distortion}\label{subsec:DeltaMag} 
Cosmic magnification causes sources---and hence, the source plane---to be magnified. The area element of the  source-plane surface (transverse to the line of sight), is given by
\begin{equation}\label{AngDist}
d\check{A} = \mathcal{\check{A}}({\bf n},z) d\Omega_{\bf n} , 
\end{equation} 
where $\check{\cal A} \,{=}\, \partial\check{A}/\partial{\Omega}_{\bf n}$ is the area per unit solid angle. (Note that the area density  $\check{\cal A}({\bf n},z) \,{=}\, \check{D}^2_A({\bf n},z)$, with $\check{D}_A$ being the well known angular diameter distance.) Generally, in an inhomogeneous universe sources may get magnified or demagnified. The amount of cosmic (de)magnification is measured by the factor $\mu$, given by \cite{Duniya:2015ths, Duniya:2016gcf}
\begin{equation}\label{Magnfcn}
\mu^{-1} \equiv \dfrac{\mathcal{\bar M}}{\check{\cal M}} = \dfrac{\check{\cal A}}{\bar{\cal A}}, 
\end{equation}
where $\check{\cal M}$ is the magnification density. (Note also that $\check{\cal F} \,{=}\, \mu \bar{\cal F}$.) The redshift-space area density contrast is given by $\check{\delta}_{\cal A} \,{\equiv}\, \delta\check{\cal A} / \bar{\cal A}$, where by a gauge transformation from real space to redshift space:  
\begin{equation}\label{perp_dA}
\check{\delta}_{\cal A}({\bf n},z) = \delta_{\cal A}({\bf n},z) -\dfrac{d\ln\bar{\cal A}}{d\bar{z}} \delta{z}({\bf n},z),
\end{equation}
with $\delta_{\cal A} \,{\equiv}\, \delta{\cal A} / \bar{\cal A}$ being the real-space area density contrast. In real-space coordinates, the infinitesimal area element is given by
\begin{eqnarray}\label{dA1}
dA &=& \sqrt{-\tilde{g}}\, \epsilon_{\mu\nu\alpha\beta}\, \tilde{u}^\mu \tilde{\ell}^\nu d\tilde{x}^\alpha d\tilde{x}^\beta,\\ \label{dA2}
&\equiv & \mathcal{A} (\theta_O,\vartheta_O)\, d\theta_O d\vartheta_O, 
\end{eqnarray}
where,
\begin{equation}\label{A:defn}
{\cal A} = \sqrt{-\tilde{g}}\, \epsilon_{\mu\nu\alpha\beta}\, \tilde{u}^\mu \tilde{\ell}^\nu \dfrac{\partial\tilde{x}^\alpha}{\partial\theta_S} \dfrac{\partial\tilde{x}^\beta}{\partial\vartheta_S} \left\| \left. \dfrac{\partial(\theta_S,\vartheta_S)}{\partial(\theta_O,\vartheta_O)} \right\| \right. ,
\end{equation}
with $\tilde{\ell}^\nu$ being a 4-vector orthogonal to the line of sight, i.e. $\tilde{u}_\nu\tilde{\ell}^\nu \,{=}\, 0$, given by \cite{Duniya:2015ths, Duniya:2016gcf, Jeong:2011as} 
\begin{equation}\label{ell:defn}
\tilde{\ell}^\nu = \tilde{u}^\nu + \dfrac{\tilde{n}^\nu}{\tilde{n}^\alpha \tilde{u}_\alpha}.
\end{equation}
After some calculations (see Appendix~\ref{appendix:DMD} for details): 
\begin{eqnarray}\label{A:exprssn}
\dfrac{\mathcal{A}}{\mathcal{\bar A}} &=& 1 - 3D - \phi + \bar{n}^i B_{\mid i} - \dfrac{1}{2} \delta{g}_{\alpha\beta}\bar{n}^\alpha \bar{n}^\beta + 2\dfrac{\delta{r}}{\bar{r}} \nonumber\\
&&+\; \left(\cot{\theta_O} +\partial_{\theta}\right)\delta{\theta} +\partial_{\vartheta}\delta{\vartheta} ,
\end{eqnarray}
where $\mathcal{\bar A}(\bar{z}) \,{=}\, a(\bar{z})^2\bar{r}(\bar{z})^2\sin{\theta_O}$. Given~\eqref{Magnfcn}, \eqref{perp_dA} and \eqref{A:exprssn}, we have the (de)magnification factor given by
\begin{align}\label{Mag2}
\mu^{-1} &= 1 - 3D - \phi + \bar{n}^i B_{\mid i} - \dfrac{1}{2} \delta{g}_{\alpha\beta}\bar{n}^\alpha \bar{n}^\beta 
+ 2\dfrac{\delta{r}}{\bar{r}} \nonumber\\
&+ \left(\cot{\theta_O} +\partial_{\theta}\right)\delta{\theta} +\partial_{\vartheta}\delta{\vartheta} + 2a\left(1 - \dfrac{1}{\bar{r} {\cal H}}\right)\delta{z},
\end{align}
where,
\begin{equation}
\dfrac{d\ln\mathcal{\bar A}}{d\ln(1+\bar{z})} = -2 + \dfrac{2}{\bar{r} {\cal H}} .
\end{equation}
By applying the appropriate equations, and after some calculations (see Appendix~\ref{appendix:DMD}), we have
\begin{eqnarray}\label{dmag1}
\mu^{-1}  &=& 1 - 2\Psi + \int^{\bar{r}_S}_0{ d\bar{r} \left[\dfrac{2}{\bar{r}_S} + \left(\bar{r} -\bar{r}_S\right) \dfrac{\bar{r}}{\bar{r}_S} \nabla^2_\perp \right] \left(\Phi +\Psi\right) } \nonumber\\
&&+\; 2\left(\dfrac{1}{\bar{r}_S{\cal H}} - 1\right) \left[\Phi + {\bf n}\, {\cdot} {\bf V} + \int^{\bar{r}_S}_0{ d\bar{r} \left(\Phi' + \Psi'\right) }\right] .
\end{eqnarray}
Thus, with $\mu^{-1} \,{=}\, 1-\check{\delta}_{\cal M}$ and, given \eqref{Phi}--\eqref{eq:33}, \eqref{delta_r2}, \eqref{dVol} and \eqref{Mag2}, we have the relativistic expression for the observed \emph{magnification overdensity}, given by
\begin{eqnarray}\label{def:MagDens}
\Delta^{\rm obs}_{\cal M}({\bf n},z) &{\equiv}& {\cal Q}(\bar{z}) \check{\delta}_{_{\cal M}}({\bf n},z), \\ \label{MagDelta}
 &=& {\cal Q}\int^{\bar{r}_S}_0{ d\bar{r}\left(\bar{r}_S - \bar{r}\right) \dfrac{\bar{r}}{\bar{r}_S} \nabla^2_\perp \Big(\Phi + \Psi\Big) ({\bf n},\bar{r})} \nonumber\\
&&+\;  2{\cal Q}\Psi({\bf n},z) - \dfrac{2{\cal Q}}{\bar{r}_S} \int^{\bar{r}_S}_0{ d\bar{r} \Big(\Phi + \Psi\Big) ({\bf n},\bar{r})} \nonumber\\
&&+\; 2{\cal Q}\left(1 -\dfrac{1}{\bar{r}_S{\cal H}}\right) \Big[\Phi({\bf n},z) + {\bf n}\, {\cdot} {\bf V}({\bf n},z) \nonumber\\
&& \hspace{1.1cm}  + \int^{\bar{r}_S}_0{d\bar{r} \Big(\Phi' + \Psi' \Big)({\bf n},\bar{r}) } \Big] .
\end{eqnarray}
The physical terms in \eqref{MagDelta} denote the same cosmological effects as in \eqref{Delta_n}, except that here they have different cofactors---which include the magnification bias. It should be pointed out that gravitational lensing is the standard source of cosmic magnification in an inhomogeneous universe. However, as given by \eqref{MagDelta} (see also e.g.~\cite{Jeong:2011as,Duniya:2016gcf}), gravitational lensing is not the only source of cosmic magnification: other sources are constituted by the relativistic effects, i.e. time delay (integral term in the second line), Doppler effect (velocity term in square brackets), ISW effect (integral term in square brackets) and, potential difference (other non-integral terms). Moreover, cosmic magnification is only one of the effects of gravitational lensing: the other effect is \emph{cosmic shear}~\cite{Bonvin:2008ni, VanWaerbeke:2009fb, Weinberg:2012es, Duncan:2013haa, Umetsu:2015baa, Gillis:2015caa}.

The cosmic magnification will be crucial in understanding cosmic distances and the nature of large-scale structure in the universe, using the data from future surveys that depend on the apparent flux and/or angular size of the sources, such as surveys of the 21 cm emission line of neutral hydrogen (HI) by the SKA~\cite{Camera:2014bwa, Maartens:2015mra}, and the baryon acoustic oscillation (BAO) surveys by BOSS~\cite{Eisenstein:2011sa, Albareti:2016xlm}. The cosmic magnification will also be key to probing the geometry of the universe.

Notice that the relativistic overdensities $\Delta^{\rm obs}_n$ and $\Delta^{\rm obs}_{\cal M}$---which are, separately, observable---are volume $\check{\upsilon}$ and flux $\check{F}$ dependent, respectively. (The methods given by e.g.~\cite{Duniya:2016gcf, Schmidt:2009rh, Gillis:2015caa, Heavens:2011ei} can be applied to measure $\Delta^{\rm obs}_{\cal M}$.) Thus, $\Delta^{\rm obs}_{\cal M}$ and $\Delta^{\rm obs}_n$ essentially comprise the ``magnified'' and the ``unmagnified'' parts, respectively, of the observed (relativistic) galaxy overdensity \eqref{ObzDelta2}.  

Thus, the general expression for the total observed galaxy overdensity \eqref{ObzDelta2}, is given by \eqref{ObzDelta4}: 
\begin{eqnarray}\label{ObzDelta3}
\Delta^{\rm obs}_{\rm g}({\bf n},z) &=& \epsilon^2_{\cal V}(z) \Delta^{\rm obs}_n({\bf n},z) + \epsilon^2_{\cal F}(z) \Delta^{\rm obs}_{\cal M}({\bf n},z),\\ \label{ObzDelta4}
&=& \epsilon^2_{\cal V}(z) \Big\lbrace \Delta_{{\rm g}}({\bf n},z) + \dfrac{1}{{\cal H}}  \partial_r \left({\bf n}\, {\cdot} {\bf V} \right)({\bf n},z) + \dfrac{1}{{\cal H}}\Psi'({\bf n},z) \nonumber\\
&&\hspace{1cm} +\; (3-b_e){\cal H}V({\bf n},z) \Big\rbrace - 2(1-{\cal Q}_{\rm eff})\Psi({\bf n},z)  \nonumber\\
&+& \left[ \epsilon^2_{\cal V}(\bar{z})\left(3 - b_e + \dfrac{{\cal H}'}{{\cal H}^2}\right)  - 2(1-{\cal Q}_{\rm eff})\left(1 - \dfrac{1}{\bar{r}_S {\cal H}}\right) \right] \Phi({\bf n},z) \nonumber\\
&+& (1-{\cal Q}_{\rm eff})\int^{\bar{r}_S}_0{d\bar{r} \left[\dfrac{2}{\bar{r}_S} - \left(\bar{r}_S - \bar{r} \right) \dfrac{\bar{r}}{\bar{r}_S } \nabla^2_\perp\right] \Big(\Phi + \Psi\Big)({\bf n},\bar{r})} \nonumber\\
&+& \left\lbrace \epsilon^2_{\cal V}(z)\left(2 - b_e + \dfrac{{\cal H}'}{{\cal H}^2}\right)  - 2(1-{\cal Q}_{\rm eff}) \left(1 - \dfrac{1}{\bar{r}_S {\cal H}}\right) \right\rbrace \nonumber\\
&&\hspace{1cm}\times \left[{\bf n}\, {\cdot} {\bf V}({\bf n},z) + \int^{\bar{r}_S}_0{d\bar{r} \Big(\Phi' + \Psi' \Big)({\bf n},\bar{r}) } \right] ,
\end{eqnarray}
where,
\begin{eqnarray}\label{Qeff}
1 - {\cal Q}_{\rm eff} \equiv \epsilon^2_{\cal V} - \epsilon^2_{\cal F} {\cal Q},
\end{eqnarray}
with ${\cal Q}$ as given by \eqref{b_M}. The observed, relativistic overdensity \eqref{ObzDelta4} is unique and physically defined: it does not correspond to any of the standard gauge-invariant definitions of overdensity, but it is automatically gauge-invariant~\citep{Duniya:2013eta}. Moreover, the comoving galaxy overdensity, the redshift-space distortion and the magnification overdensity, give galaxy surveys the potential to probe the content and the geometry of the universe~\cite{Bonvin:2014owa}. 

The parameters $\epsilon_{\cal V}$ and $\epsilon_{\cal F}$ have never previously been taken into account in the literature, in the calculation of the observed, relativistic galaxy overdensity \eqref{ObzDelta3}. The calculations in \eqref{vect-dN1}--\eqref{ObzDelta4} therefore outline---for the first time---a clear, consistent approach for combining the ``magnified'' (flux-dependent) and the ``unmagnified'' (volume-dependent) relativistic components of the observed overdensity of galaxy surveys. 

Equation \eqref{ObzDelta4} generalizes the observed galaxy overdensity, correct for a generic survey (or sample). It applies to both (i) surveys that depend on magnification ($\epsilon_{\cal F} \,{\neq}\, 0$), e.g. the HI surveys of the SKA~\cite{Camera:2014bwa, Maartens:2015mra} and the BAO surveys of BOSS~\cite{Eisenstein:2011sa, Albareti:2016xlm}, both which depend on the apparent flux and/or angular size of the sources; and (ii) surveys that are independent of magnification ($\epsilon_{\cal F} \,{=}\, 0$), e.g. the galaxy surveys of DES~\cite{Diehl:2014lea} and EUCLID~\cite{Scaramella:2015rra}, whose selection probability depends only on the intrinsic (physical) properties, the redshifts and the observed sky positions of the sources. (See section \ref{sec:FVLS} for discussion on specific sample types.)

\subsection{The Linear Galaxy Bias}\label{subsec:b_g}
The observation of galaxies is based on measuring emitted photons from the galaxies. This provides information of the distribution and the nature of the given galaxies; hence, probing the actual mass over the given cosmic scales. However, the relation between the distribution of the galaxies and the distribution of the associated mass is not one-to-one: the photons from these galaxies do not accurately trace the underlying matter content. The inability of the galaxies to map the true underlying mass is corrected by a factor, called the \emph{galaxy bias} \cite{Bartolo:2010ec, Baldauf:2011bh, Jeong:2011as, Duniya:2015ths}. It captures the uncertainty between the galaxy distribution and the underlying mass. In linear perturbations, the galaxy bias is sufficiently described by time dependence only: it is purely a function of time. 

As shown by \cite{Duniya:2015ths, Duniya:2015dpa}, by taking a gauge transformation from an arbitrary coordinate frame into the matter (``$m$'') rest frame (``rf''), we have
\begin{equation}\label{rf-drho}
\left. \dfrac{\delta{\rho}_m}{\bar{\rho}_m} \right\|_{\rm rf} = \dfrac{\delta{\rho}_m}{\bar{\rho}_m} + \left(v_m+B\right)\dfrac{\bar{\rho}_m'}{\bar{\rho}_m} \;\equiv\; \Delta_m, 
\end{equation}
where $\bar{\rho}_m$ and $\delta{\rho}_m$ are the background and the coordinate (energy) density perturbation, respectively; $v_m$ is the scalar potential of the coordinate peculiar velocity; $B$ is a scalar perturbation of the metric \eqref{conformal-ds2} and, $\Delta_m$ is the (gauge-invariant) comoving overdensity. Similarly, 
\begin{equation}\label{rf-dg_n}
\left. \dfrac{\delta{n}_{\rm g}}{\bar{n}_{\rm g}} \right\|_{\rm rf} = \Delta_{\rm g}, 
\end{equation}
where the comoving galaxy (number) overdensity $\Delta_{\rm g}$, is as given in \eqref{flatCurvDelta}. Thus, the comoving galaxy overdensity is related to the comoving matter overdensity, given by 
\begin{equation}\label{galaxy-matter}
\Delta_{\rm g} = b\/\Delta_m,
\end{equation} 
where $b \,{=}\, b(a)$ is the galaxy bias, given by
\begin{equation}\label{b_g}
b \;\equiv \left. \dfrac{\partial\ln(\bar{n}_{\rm g})}{\partial\ln(\bar{\rho}_m)} \right\|_a ,
\end{equation}
which is sufficient for linear perturbations. Thus, given \eqref{b_e}, \eqref{rf-drho}, \eqref{rf-dg_n} and \eqref{b_g}, we relate the galaxy bias to the (galaxy) evolution bias, given by 
\begin{equation}\label{galbias}
b = -{\cal H} \left(3 -b_e\right) \dfrac{\bar{\rho}_m}{\bar{\rho}'_m} = 1 - \dfrac{1}{3}b_e,
\end{equation}
where we used the matter background conservation equation: $\bar{\rho}'_m \,{=}\, {-}3{\cal H}\bar{\rho}_m$ (for non-interacting dark sector).

In Ref.~\cite{Jeong:2011as}, the authors used a different approach to derive a relation between $b$ and $b_e$. In the given reference, the authors assumed that the abundance of galaxies follows a universal mass function; whence they work to obtain $b \,{=}\, 1 \,{+}\, b_e/(\delta_c f)$, with $\delta_c$ being the matter collapse density contrast and, $f$ being the usual logarithmic matter growth rate of the standard concordance model, $\Lambda$CDM: a universe dominated by a cosmological constant $\Lambda$ and cold dark matter (CDM). (See \cite{Duniya:2015ths, Duniya:2015nva} for the definition of $f$ that is correct for all models.) It should be noted that in \eqref{galbias} we did not use any assumptions in the derivation: only the natural definition of the parameters.


\section{Flux and Volume Limited Samples}\label{sec:FVLS}
A generic survey would involve both, varying apparent flux (with various apparent magnitudes continuously measured) and a varying survey volume (with various redshifts continuously measured), simultaneously. The generated catalogue inherently contains both a ``magnified'' (flux-dependent) and an ``unmagnified'' (volume-dependent) fractions, respectively, with the measurable quantity---being the number of galaxies per unit solid angle per redshift---comprising both fractions together, without any distinctions. 

In practice, however, analyses are often done for specific samples: usually, for either flux-limited or volume-limited samples. In order to construct or extract these samples, certain measurement limits are defined.

\subsection{The Limits} 
The principal limiting parameters are usually the apparent flux and the observed (geometric) cosmic depth. These parameters invariably determine the source intrinsic brightness (luminosity). The flux is prescribed by an apparent magnitude $m$ and, the luminosity is given by an absolute magnitude $M \,{<}\, 0$. 

Once the apparent magnitude limit $m_{\rm lim}$ is specified, the limit on the absolute magnitude is determined (via the magnitude-distance relation) by
\begin{eqnarray}\label{M_lim}
M_{\rm lim} = m_{\rm lim} - 25 - 5\log_{10} (D_{\rm lim}),
\end{eqnarray}
where $D_{\rm lim} \,{=}\, \bar{r}(z_{\rm lim})$ is the cosmic depth, being the background distance (in ${\rm Mpc}$) along the line of sight, at a given maximum redshift value $z_{\rm lim}$. Note that in reality, $M_{\rm lim}$ is affected by the well-known $K$-correction, galactic extinction, and other notable effects---especially if $D_{\rm lim}$ is large (i.e.~at high $z$)---and thus, a more general expression for $M_{\rm lim}$ will need to incorporate these effects~\citep{Martinez:2002bk}.

\subsection{Flux-limited Samples}\label{subsec:FLS}
A flux-limited sample may be obtained directly from experiment, by a flux-limited survey---in which only sources of the same apparent magnitude (above the instrument threshold) are allowed, at varying redshift. The instrument needs to have the sufficient resolution and sensitivity to determine $z$. This way the individual sources are differentiated, and their actual number is counted. (Invariably, a flux-limited survey is also a ``redshift'' survey, with the catalogue consisting of sky positions and luminosities of the sources.) 

Flux-limited samples may also be obtained from a generic galaxy catalogue data: a flux value of choice is adopted and set as the working limit for the required selection and, only sources with an apparent flux which is equal to the working limit are extracted from the catalogue. The adopted flux limit is fixed by the apparent magnitude value $m_{\rm lim}$ and, sources whose apparent brightness is equal to this limit are selected---at various cosmic distances (or redshifts)---given by
\begin{eqnarray}\label{m_lim}
M = m_{\rm lim} - 25 - 5\log_{10} (\bar{r}),
\end{eqnarray}
where $m_* \,{=}\, m_{\rm lim}$. We see that the absolute magnitude is mainly a function of the background (line-of-sight) comoving distance $\bar{r}$. 

Note that for a flux-limited sample, by having the same apparent magnitude, it implies that sources at higher redshifts must have higher luminosities (intrinsic brightness) in order to be detected by the instrument, i.e. sources at larger $\bar{r}$ must have lower $M$. In other words, since the resulting distribution density depends on the sources' distance---and independent of the apparent magnitude---it implies that intrinsically faint sources ($M \,{>}\, M_{\rm lim}$) will not be detected by the instrument even at a small distance, unless they are close enough (to the Earth); whereas, brighter sources ($M \,{\leq}\, M_{\rm lim}$) are detected even at a large distance away. This creates a gradient in the number density: $\partial\bar{n}_{\rm g}/\partial\bar{r} \,{\neq}\, 0$ (with $\partial\bar{n}_{\rm g}/\partial{m} \,{=}\, 0$). Thus, the given sample is spatially non-uniform \citep{Martinez:2002bk}.

Given that the flux-limited sample has a constant apparent magnitude, there is no change in flux, i.e. $d\check{F} \,{=}\, 0$ and, the flux per unit solid angle per redshift, $\bar{\cal F} \bar{q}_{\cal X}$, vanishes. Consequently, $\epsilon_{\cal F} \,{=}\, 0$ and by \eqref{fracSum} we have $\epsilon_{\cal V} \,{=}\, 1$, with the average distribution density being given by $\check{\left< {\cal N}_{\rm g} \right>} \,{=}\, \bar{n}_{\rm g}\bar{\cal V}$. Then by \eqref{ObzDelta3}, the given sample is fully described by the number-count overdensity \eqref{Delta_n}. (Appropriate for pure number-count analysis---see e.g.~\cite{Bonvin:2011bg, LopezHonorez:2011cy, Durrer:2016jzq}.)

\subsection{Volume-limited Samples}\label{subsec:VLS} 
An ideal volume-limited survey would be one done by observing a fixed patch of the sky at a specified redshift, and the flux and/or size distribution on the given sky patch is mapped---i.e. only sources at a given (fixed) redshift are observed. The sample will contain sources of different apparent magnitudes (above the instrument threshold), but all having the same $z$ measurement. 

However, volume-limited samples may also be extracted from generic catalogues of galaxy surveys. This is usually done by specifying a depth $D_{\rm lim}$---being the maximum cosmic reach from the observer---and picking out only sources at that given depth, with luminosities specified by an absolute magnitude, given by
\begin{eqnarray}\label{D_lim}
M = m - 25 - 5\log_{10} (D_{\rm lim}),
\end{eqnarray}
where ${r}_* \,{=}\, D_{\rm lim}$ and, we see that the absolute magnitude is mainly a function of the apparent magnitude $m$, and hence, of the flux (density). Here, the sources are accessed by varying the apparent magnitude. Moreover, the cosmic depth $D_{\rm lim}$ is essentially the limiting radius of the cosmic sample. We notice that, given \eqref{D_lim}, in volume-limited samples the flux limit is no longer present and, all possible fluxes are mapped (at the fixed cosmic distance). The extracted sample is spatially uniform---in the sense that the gradient of the number density vanishes, $\partial\bar{n}_F/\partial\bar{r} \,{=}\, 0$ (with $\partial\bar{n}_F/\partial{m} \,{\neq}\, 0$): the number density of the given sample does not vary with distance~\citep{Martinez:2002bk}.

In general, for volume-limited samples there is no change in volume, i.e. $d\check{\upsilon} \,{=}\, 0$ and, the volume per unit solid angle per redshift, $\bar{\cal V}$, vanishes. Consequently, $\epsilon_{\cal V} \,{=}\, 0$ and,  by \eqref{fracSum} we have $\epsilon_{\cal F} \,{=}\, 1$, with the average distribution density being given by $\check{\left< {\cal N}_{\rm g} \right>} \,{=}\, \bar{n}_F \bar{\cal F} \bar{q}_{\cal X}$. Thus, the generated sample is described by the magnification overdensity \eqref{MagDelta}. (Such kind of samples are appropriate for pure magnification analysis---see e.g. \cite{Gillis:2015caa, Duniya:2016gcf}.)

In previous analyses (see e.g. \cite{Challinor:2011bk, Jeong:2011as, LopezHonorez:2011cy, Duniya:2015ths, Alonso:2015uua}), in order to compute predictions for galaxy number counts, the magnification bias is set to zero (${\cal Q} \,{=}\, 0$), which is ad hoc. However, in this work, the total observed overdensity \eqref{ObzDelta4} does not require ${\cal Q} \,{=}\, 0$ or the absolute absence of magnification in the background distribution of sources, but as explained above, it is sufficient to assume the physical limiting condition applied when extracting flux-limited samples from a catalogue, which is to fix the observed magnitude at a certain brightness value and intrinsically bright sources of absolute magnitude within a given range are selected into the sample (noting that the given absolute magnitude is only dependent on cosmic distance). Consequently, there is no change in flux, which leads to the vanishing of the fraction $\epsilon_{\cal F}$ of intrinsically faint sources in the background distribution density; thereby eliminating the effect of cosmic magnification and the magnification overdensity in the total observed overdensity (with $\epsilon_{\cal F} \,{=}\, 0$). Physically, this makes sense since the detection of cosmic magnification depends on variation in flux (or angular size), hence any possible magnified (faint) sources will not be selected given that the value of apparent magnitude for detection is the same for all sources. (This can not be explained by merely setting ${\cal Q} \,{=}\, 0$.)

On the other hand, for volume-limited samples---which are used in the analysis of cosmic magnification---the calculation of the total observed overdensity in previous works in the literature will fail to reduce to the magnification overdensity \eqref{MagDelta}, as there is no longer any parameters to apply any limiting condition on. This is not the case for the calculation of the total observed overdensity \eqref{ObzDelta4} in this work. In order to extract volume-limited samples from survey catalogues, it is common practice that astronomers usually specify a given fixed (geometric) depth, and then select only sources whose cosmic distance is at the given depth. This corresponds to observing sources in a fixed volume, with the radius of the volume defined by the fixed depth. Consequently, there is not change in volume, which leads to the vanishing of the fraction $\epsilon_{\cal V}$ of intrinsically bright sources in the background distribution density; thereby eliminating the effect of volume (or distance) variation and the overdensity of number counts in the total observed overdensity (with $\epsilon_{\cal V} \,{=}\, 0$). Physically, this makes sense since detecting intrinsically bright sources essentially involves increasing the redshift reach of the instrument---which will imply encompassing larger cosmic distances and hence, observed volume---hence by fixing the cosmic depth, the distribution of the observed sources must vary only with apparent magnitude. Thus, in this case the total observed overdensity reduces to the relativistic magnification overdensity. (Again, this can not be shown with the result in previous related works.)


\section{The Observed Angular Power Spectrum}\label{sec:OAPS}
We investigate the angular power spectrum associated with \eqref{ObzDelta3}. We avoided using any approximations, e.g.~the Limber's approxiamtion~\citep{LoVerde:2008re} (for integral terms). Moreover, to probe the galaxy angular power spectrum, we assume the late-time $\Lambda$CDM universe; with all the cosmic species having vanishing anisotropic stress: thus, $\Psi \,{=}\, \Phi$.

By using \eqref{ObzDelta4}, we have the observed angular galaxy power spectrum (dropping all over-bars), given by
\begin{align}\label{Cl-obs}
C^{\rm obs}_\ell(z_S) &= \left\langle{ \Big\|a_{\ell m}(z_S)\Big\|^2 }\right\rangle,\nonumber\\
&=  \dfrac{4}{\pi^2}\left(\dfrac{9}{10}\right)^2\int{dk\, k^2 T(k)^2 P_{\Phi_p}(k) \Big\| f_\ell(k,z_S) \Big\|^2 },
\end{align}
where $T(k)$ is the linear transfer function and $P_{\Phi_p}(k)$ is the primordial power spectrum~\citep{Dodelson:2003bkk}; $f_\ell$ is given by \eqref{f_ell}:
\begin{eqnarray}\label{f_ell}
f_\ell(k,z_S) &=& \epsilon^2_{\cal V} \left\lbrace \left[ b \widehat{\Delta}_m(k,z_S) + \left(3 - b_e\right){\cal H}\widehat{V}_m(k,z_S) + \dfrac{1}{{\cal H}}\widehat{\Phi}'(k,z_S) \right] j_\ell(kr_S) \right.\nonumber\\
&&\hspace{1cm} \left. -\; \dfrac{1}{{\cal H}} \partial_r \widehat{V}^\parallel_m(k,z_S) \partial^2_{kr}j_\ell(kr_S) \right\rbrace \nonumber\\
&+& \left[ \epsilon^2_{\cal V} \left(3 - b_e + \dfrac{{\cal H}'}{{\cal H}^2}\right)  - 4(1 - {\cal Q}_{\rm eff}) + \dfrac{2(1 - {\cal Q}_{\rm eff})}{r_S {\cal H}} \right] \widehat{\Phi}(k,z_S) j_\ell(kr_S) \nonumber\\
&+& \left\lbrace \epsilon^2_{\cal V} \left(b_e - 2 - \dfrac{{\cal H}'}{{\cal H}^2}\right) + 2(1-{\cal Q}_{\rm eff}) \left(1 - \dfrac{1}{r_S {\cal H}}\right) \right\rbrace \nonumber\\
&& \hspace{0.5cm} \times\;  \left[ \widehat{V}^\parallel_m(k,z_S) \partial^{ }_{kr} j_\ell(kr_S) - 2\int^{r_S}_0{dr\, j_\ell(kr) \widehat{\Phi}'(k,r) } \right] \nonumber\\
&+& 2(1-{\cal Q}_{\rm eff}) \int^{r_S}_0{dr\, j_\ell(kr) } \left[\dfrac{2}{r_S} - \dfrac{(r {-} r_S)}{rr_S} \ell\left(1+\ell\right) \right] \widehat{\Phi}(k,r) ,
\end{eqnarray}
where $V_\parallel \,{\equiv}\, {-}{\bf n}\, {\cdot} {\bf V} \,{=}\, \partial_r V$ gives the line-of-sight component of the peculiar velocity, and ${\cal Q}_{\rm eff}$ is as given by \eqref{Qeff}, $j_\ell$ is the spherical Bessel function, $\partial_{kr} \,{=}\, \partial/\partial(kr)$ and, $\widehat{X}(k,z) \,{\equiv}\, X(k,z)/\Phi_d(k)$ for a given parameter $X$; $\Phi_d$ is the gravitational potential at the photon-matter decoupling $z\,{=}\,z_d$, with~\citep{Duniya:2013eta, Duniya:2015nva, Duniya:2015ths, Dodelson:2003bkk}
\begin{eqnarray}\label{Phi_d}
\Phi(k,z_d) = \dfrac{9}{10} \Phi_p(k) T(k) \equiv \Phi_d(k),
\end{eqnarray}
where $\Phi_p$ is the primordial gravitational potential. (See \cite{Dodelson:2003bkk} for a numerical formula for $T(k)$.) We have also used the fact that on very large (linear) scales, which are the scales under consideration in this work, the matter and the galaxy rest frames coincide (given the homogeneity and isotropy of these scales), and we set the matter and the galaxy velocities---along the line of sight---to be equal: $V^\parallel_m \,{=}\, V_\parallel \,{=}\, V^\parallel_{\rm g}$. In other words, galaxies follow trajectories that are parallel to that of the underlying matter (equivalently, the galaxies are comoving with the underlying matter); hence there is essentially no momentum (ex)change between the two frames to cause any bias between the velocities \cite{Duniya:2015ths}. 

For all numerical computations we use: a present-epoch matter density parameter $\Omega_{m0} \,{=}\, 0.3$; the Hubble constant $H_0 \,{=}\, 67.8$~km$\cdot$s$^{-1}\cdot$Mpc$^{-1}$; an evolution bias $b_e \,{=}\, 0$---a scenario where the galaxies do not evolve with time (the background galaxy comoving number density is constant)---which given \eqref{galbias}, the galaxy bias $b \,{=}\, 1$. Although the analysis of the observed overdensity \eqref{ObzDelta4} needs to be done for specific sample types, here we give the angular power spectrum only for illustrative purpose; $\epsilon_{\cal V}$ and $\epsilon_{\cal F}$ are taken as constants.

\begin{figure*}\centering
\includegraphics[width=12cm,height=8cm]{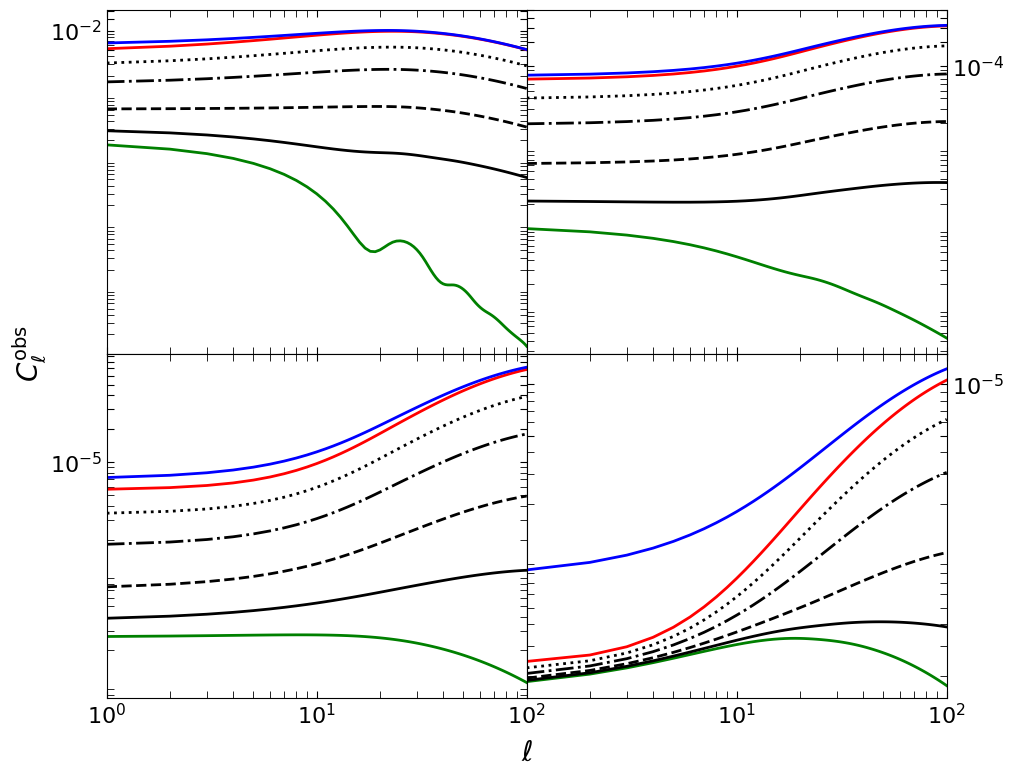} 
\caption{The plots of the angular power spectra for magnification bias ${\cal Q} \,{=}\, 1$, at the source redshifts: $z_S \,{=}\, 0.1$ ({\it top left}), $z_S \,{=}\, 0.5$ ({\it top right}), $z_S \,{=}\, 1$ ({\it bottom left}) and $z_S \,{=}\, 3$ ({\it bottom right}). The various lines show the full relativistic angular power spectrum $C^{\rm obs}_\ell$, as a function of the multipole order $\ell$, for: $\epsilon_{\cal V} \,{=}\, 0$ and $\epsilon_{\cal F} \,{=}\, 1$ (green), which gives the magnification angular power spectrum; $\epsilon^2_{\cal V} \,{=}\, 0.1$ and $\epsilon^2_{\cal F} \,{=}\, 0.9$ (solid black); $\epsilon^2_{\cal V} \,{=}\, 0.25$ and $\epsilon^2_{\cal F} \,{=}\, 0.75$ (dashed black); $\epsilon^2_{\cal V} \,{=}\, 0.5 \,{=}\,\epsilon^2_{\cal F}$ (dot-dashed black), $\epsilon^2_{\cal V} \,{=}\, 0.75$ and $\epsilon^2_{\cal F} \,{=}\, 0.25$ (dotted black) and, $\epsilon_{\cal V} \,{=}\, 1$ and $\epsilon_{\cal F} \,{=}\, 0$ (red)---which gives the number-count power spectrum. The blue line shows the result corresponding to previous works ($\epsilon^2_{\cal V} \,{=}\, 1 \,{=}\, \epsilon^2_{\cal F}$).}\label{fig:Cls_bM_1}
\end{figure*}

In Fig.~\ref{fig:Cls_bM_1} we show plots of the total observed angular power spectrum $C^{\rm obs}_\ell$, the observed number-count angular power spectrum $C^{(n)}_\ell$ and the observed magnification angular power spectrum $C^{(\cal M)}_\ell$, at source redshifts $z_S \,{=}\, 0.1,\, 0.5,\, 1$ and $3$; with magnification bias ${\cal Q} \,{=}\, 1$. We show $C^{\rm obs}_\ell $ for different values of $\epsilon_{_{\cal V}}$ and $\epsilon_{\cal F}$ such that \eqref{fracSum} holds, and also for when both parameters are unity (which corresponds to the results of previous works in the literature). We see that, for all the cases where \eqref{fracSum} holds true, the total observed angular power spectrum is bounded by $C^{(\cal M)}_\ell \leq C^{\rm obs}_\ell \leq C^{(n)}_\ell$ on all scales, at all the redshifts (for the given value of ${\cal Q}$). Moreover, the amplitude of $C^{\rm obs}_\ell$ increases as the value of $\epsilon_{\cal V}$ increases: $0 \,{\leq}\, \epsilon^2_{\cal V} \,{\leq}\, 1$ with $\epsilon^2_{\cal F} \,{=}\, 1 \,{-}\, \epsilon^2_{\cal V}$. The amplitude of $C^{\rm obs}_\ell$ where $\epsilon_{\cal V} \,{=}\, \epsilon_{\cal F} \,{=}\, 1$---which corresponds to previous works---is, as should be expected, higher on ultra-large scales than when $\epsilon_{\cal V}$ and $\epsilon_{\cal F}$ are governed by \eqref{fracSum}. This difference in amplitude increases with increase in redshift, particularly at $z \,{\geq}\, 1$. In general, the results show that by computing the ``magnified'' and the ``unmagnified'' components of the observed, relativistic overdensity and then adding these components directly, the resulting overdensity can lead to an overestimation of the observed angular power spectrum on ultra-large scales. This may result in a false prediction of the large-scale clustering of galaxies. Moreover, as further shown by the results, the amount by which the large-scale clustering of galaxies is over-projected increases with increasing redshift $z \,{\geq}\, 1$ (for the given value of ${\cal Q}$).

\begin{figure*}\centering
\includegraphics[width=12cm,height=8cm]{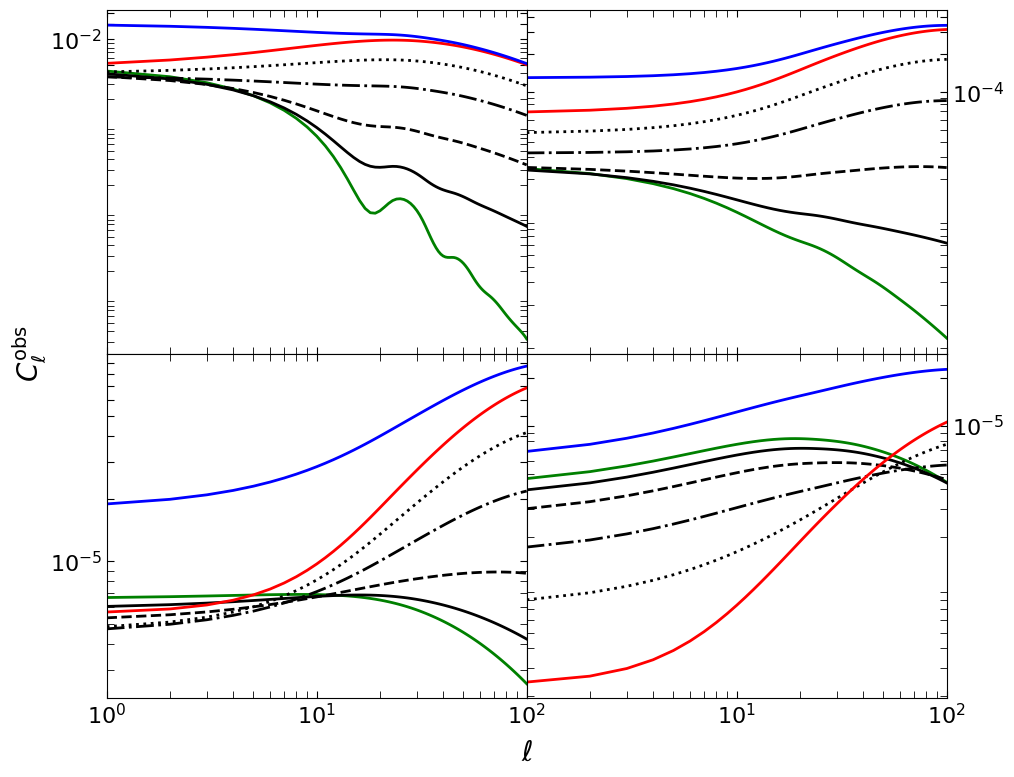} 
\caption{The plots of the angular power spectra for magnification bias ${\cal Q} \,{=}\, 5$, at the source redshifts: $z_S \,{=}\, 0.1$ ({\it top left}), $z_S \,{=}\, 0.5$ ({\it top right}), $z_S \,{=}\, 1$ ({\it bottom left}) and $z_S \,{=}\, 3$ ({\it bottom right}). The line notations are as in Fig.~\ref{fig:Cls_bM_1}. The results for ${\cal Q} \,{=}\, {-}1$ and ${\cal Q} \,{=}\, {-}5$ are identical to those of ${\cal Q} \,{=}\, 1$ and ${\cal Q} \,{=}\, 5$, respectively. Hence those results are not shown in this work.}\label{fig:Cls_bM_5}
\end{figure*}

In Fig.~\ref{fig:Cls_bM_5} we illustrate the effect of a large magnification bias. We repeated the computations of Fig.~\ref{fig:Cls_bM_1}, except that here we used the value of magnification bias ${\cal Q} \,{=}\, 5$. We see that most of the features in Fig.~\ref{fig:Cls_bM_1} are retained in these results. However, we observe reduction in the spread of the lines---and the eventual crossings, at higher redshifts $z \,{\geq}\, 1$---for multipoles $\ell \,{\lesssim}\, 20$. As already known, we see that the amplitude of the galaxy power spectrum decreases with increasing redshift. Moreover, the dominant terms in the total observed galaxy overdensity \eqref{ObzDelta3} come from the observed number-count overdensity \eqref{Delta_n} at most $z$. We see that the main effect of the higher value of ${\cal Q}$ is the boosting of the angular power spectrum, for increasing values of $\epsilon_{\cal F}$. It should be pointed out that when the magnification bias ${\cal Q} \,{=}\, 1$, there are no any terms in the observed galaxy overdensity \eqref{ObzDelta3} that will vanish (unless $\epsilon^2_{\cal V} \,{=}\, \epsilon^2_{\cal F} \,{=}\, 0.5$), unlike in the previous works in the literature (corresponding to $\epsilon_{\cal V} \,{=}\, \epsilon_{\cal F} \,{=}\, 1$) where the magnification effect cancels out almost all volume distortions. As a result, the case where ${\cal Q} \,{=}\, 1$ has been referred to as ``diffuse backgrounds'' \citep{Jeong:2011as} and ``HI intensity mapping'' \citep{Duniya:2013eta}. In this work, the so-called {\it diffuse backgrounds} or {\it HI intensity mapping} scenario, will correspond to
\begin{eqnarray}\label{HIcase}
\epsilon_{\cal V} = \sqrt{{\cal Q}}\,\epsilon_{\cal F},
\end{eqnarray}
where provided the magnification bias is positively valued, the actual value itself is irrelevant: it may or may not be unity---as enforced by previous works. (Obviously, by assuming $\epsilon_{\cal V} \,{=}\, \epsilon_{\cal F}$ in \eqref{HIcase}, we recover the given scenario of previous works, i.e.~${\cal Q} \,{=}\, 1$.) By combining \eqref{HIcase} and \eqref{fracSum}, it implies that for a so-called diffuse backgrounds or HI intensity mapping scenario, we have  $\epsilon^2_{\cal V} \,{=}\, {\cal Q}(1+{\cal Q})^{-1}$ and $\epsilon^2_{\cal F} \,{=}\, (1+{\cal Q})^{-1}$, where the magnification bias takes arbitrary values (${\cal Q} \,{\neq}\, {-}1$). Thus, the magnification effect will cancel nearly all of the volume distortion terms in the observed galaxy overdensity \eqref{ObzDelta4} when \eqref{HIcase} holds. In particular, the lensing and the time-delay terms, respectively---which constitute the dominant contribution from the volume distortion---are erased completely; whereas the remaining (subdominant) terms, being the Doppler, the ISW and the potential (difference) terms, respectively, are diminished.


\section{Conclusion}\label{sec:Concl}
A consistent, straightforward analysis of the total (observed ) relativistic overdensity of cosmological surveys was presented. Along with the already known terms, new background parameters $\epsilon_{\cal V}$ and $\epsilon_{\cal F}$ have been uncovered. The new parameters separately quantify: (i) the fraction $\epsilon_{\cal F}$ of intrinsically faint background sources which are magnified into the observed sample, by the amplification of their flux (or angular size), and (ii) the fraction $\epsilon_{\cal V}$ of the background sources which are intrinsically bright enough to be observed. These parameters dictate the contributions from the ``magnified'' (flux-dependent) and the ``unmagnified'' (volume-dependent) parts, respectively, of the total observed overdensity. The calculations show that, contrary to previous analyses, the magnified and the unmagnified components of the total observed overdensity of a generic survey do not merely add: they are scaled by$\epsilon^2_{\cal F}$ and $\epsilon^2_{\cal V}$.

Unlike in previous works in the literature---where in order to compute predictions for galaxy number counts, the magnification bias is merely set to zero (${\cal Q} \,{=}\, 0$)---in this work, the calculation of the total relativistic overdensity \eqref{ObzDelta4} does not require ${\cal Q} \,{=}\, 0$ or the absolute absence of magnification in the background distribution of sources. One only needs to invoke the same physical limiting condition applied when extracting flux-limited samples from a catalogue, which is to fix the apparent magnitude at a certain brightness value, and intrinsically bright sources of absolute magnitude within a given range are selected. In which case the absolute magnitude of the sources is only a function of cosmic distance. Consequently, there is no change in flux, which leads to the vanishing of the fraction $\epsilon_{\cal F}$ of intrinsically faint sources in the background distribution density; thereby eliminating the effect of cosmic magnification and the magnification overdensity in the total relativistic overdensity by $\epsilon_{\cal F} \,{=}\, 0$. Physically, this makes sense since the detection of cosmic magnification depends on variation in flux (or angular size); hence any possible magnified sources will not be selected given that the value of apparent magnitude for detection is the same for all sources. (This can not be explained by merely setting ${\cal Q} \,{=}\, 0$.)

Moreover, for volume-limited samples---which are used in the analysis of cosmic magnification---the calculation of the total relativistic overdensity in previous works in the literature will fail to reduce to the magnification overdensity \eqref{MagDelta}. However, this is not the case for the calculation of the total relativistic overdensity \eqref{ObzDelta4} in this work. In order to extract volume-limited samples from survey catalogues, astronomers usually specify a given fixed (geometric) depth, and then select only sources whose cosmic distance is at the given depth. Thus, the fixed depth prescribes a constant volume. Consequently, the change in volume is zero, leading to the vanishing of the fraction $\epsilon_{\cal V}$ of intrinsically bright sources in the background distribution density. This eliminates the effect of volume (or distance) variation and the overdensity of number counts by $\epsilon_{\cal V} \,{=}\, 0$. Since detecting intrinsically bright sources essentially involves increasing the redshift reach of the experiment---which implies encompassing larger cosmic distances and hence, observed volume---then by fixing the cosmic depth, the distribution of the observed sources must vary only with apparent magnitude. Thus, in this case the total relativistic overdensity reduces to the relativistic magnification overdensity. (Again, this can not be shown with the result in previous related works.)

In general, the results show that care must taken when applying the total relativistic overdensity in any cosmological analysis. Moreover, the calculations in this work serve to generalise the expression of the observed, relativistic overdensity of galaxy surveys.

\backmatter

\bmhead{Acknowledgments}
Thanks to Daniele Bertacca for useful comments. This work was carried out with financial support from (i) the government of Canada's International Development Research Centre (IDRC), and within the framework of the AIMS Research for Africa Project, and (ii) the South African Square Kilometre Array Project and the South African National Research Foundation.


\begin{appendices}

\section{The Spacetime Metric}\label{sec:metric}
Throughout this work, a flat-space Friedmann-Robertson-Walker (FRW) universe is assumed---with most of the calculations being drawn from the rigorous work by \cite{Duniya:2015ths, Duniya:2016gcf} (and relevant references therein).

The spacetime is described by a line element called the \emph{metric}~\cite{Duniya:2015ths, Durrer:1994zza, Durrer2008, HLP.TKP1994, CarmeliM:1982, Dodelson:2003bkk}, which measures the interval or distance between any two neighbouring points separated only by an infinitesimal displacement, given by 
\begin{equation}\label{AppMetric}
d\tilde{s}^2 = \tilde{g}_{\mu\nu}d\tilde{x}^\mu d\tilde{x}^\nu , 
\end{equation}
where $\tilde{g}_{\mu\nu}$ is the metric tensor, and $\tilde{x}^\mu$ denote the spacetime coordinates. The metric tensor is a symmetric tensor function of the spacetime coordinates. The metric tensor components govern matter distribution and motion through spacetime, and in turn, the motion and distribution of matter determine the metric tensor via the gravitational field equations~\cite{CarmeliM:1982, Duniya:2015ths}. 

In a perturbed FRW universe, the metric tensor may be decomposed by 
\begin{equation}\label{MetricTens}
\tilde{g}_{\mu\nu} = a^2\left(\bar{g}_{\mu\nu} + \delta g_{\mu\nu}\right),
\end{equation}
where $\bar{g}_{\mu\nu} \,{=}\, \bar{g}_{\mu\nu}(\bar \eta)$ and $\delta g_{\mu\nu} \,{=}\, \delta g_{\mu\nu}(\eta,x^i)$, denoting background and perturbation, respectively. We adopt conformal time $\eta$, where $dt \,{=}\, ad\eta$, with $t$ being physical (cosmic) time and $a \,{=}\, a(\eta)$ is the `scale factor' which measures the magnitude of the cosmic expansion; ${x}^{i}$ denotes the space $3$-vector, and
\begin{equation}\label{BKD:MetricTens}
\bar{g}_{00} = -1,\quad \bar{g}_{i0} = \vec{0} = \bar{g}_{0j},\quad \bar{g}_{ij} = \delta_{ij},
\end{equation}
where we consider (henceforth) only linear perturbations, in a flat FRW universe. Moreover, we consider only scalar perturbations, since tensor modes give rise to gravitational waves which do not interact with neither energy density nor pressure fluctuations and, any primordial vector modes would have decayed out kinematically at late times in an expanding universe \cite{Duniya:2015ths, Mukhanov:1990me}.

Thus, the perturbation $\delta g_{\mu\nu}$ may be parametrised by scalar fields: $\phi \,{=}\, \phi(\eta ,{x}^{i})$, $B \,{=}\, B(\eta ,{x}^{i})$, $D \,{=}\, D(\eta ,{x}^{i})$ and $E \,{=}\, E(\eta ,{x}^{i})$, given by
\begin{eqnarray}\label{PertMetricComps}
\delta{g}_{00} = -2\phi,~~\delta{g}_{i0} = B_{i},~~\delta{g}_{ij} = -2\left(D\delta_{ij}-E_{ij}\right),
\end{eqnarray}
where $B_{i} \,{=}\, B_{\mid  i}$ and $E_{ij} \,{\equiv}\, E_{\mid  ij} {-} \frac{1}{3}\delta_{ij}\nabla^2E$ is a traceless transverse tensor. Thus, the metric \eqref{AppMetric} in a perturbed FRW universe is given completely by
\begin{eqnarray}\label{Conf:metric}\nonumber
d\tilde{s}^2 &=& a(\eta)^2 \Big\lbrace -\left(1+2\phi\right) d{\eta}^2 + 2B_{\mid  i}\, d{\eta} dx^i  + \left[ (1 - 2\psi)\delta_{ij} +2E_{\mid  ij}\right] dx^i dx^j \Big\rbrace ,
\end{eqnarray} 
where $\psi \equiv D +\frac{1}{3}\nabla^2E$. The various given scalar-field parametrizations \eqref{PertMetricComps} of the metric perturbations completely exhaust the (scalar) perturbative degrees of freedom of the metric. The 4-velocity of a particle moving in a perturbed FRW universe, is given by \eqref{vels}.

The metric~\eqref{Conf:metric} has two displeasing features, being (i) even by using only scalar perturbations in the metric, the resulting cosmological equations remain complicated; (ii) such generic perturbations \eqref{PertMetricComps} do give rise to ghost scalar and vector modes in the solutions of these equations: the latter difficulty can be removed while the former is alleviated if one fixes the coordinate system~\cite{Weinberg:2008}, or by using gauge-invariant quantities. 

Given \eqref{Conf:metric} and \eqref{vels}, we define gauge-invariant potentials, given by
\begin{eqnarray}\label{InvPhi}
\Phi &{\equiv}& \phi -{\cal H} \sigma -\sigma',\\ \label{InvPsi}
\Psi &{\equiv}& \psi + {\cal H} \sigma, \\ \label{InvVel}
V &{\equiv}& v + E',
\end{eqnarray}
where $\sigma = {-}B+E'$ and ${\cal H} \,{=}\, a'/a$ is the comoving Hubble parameter, and a prime denotes derivative with respect to conformal time.


\section{The Distortions in the Observed Galaxy Overdensity}\label{appendix:DMD}
The calculations in this appendix are drawn from the rigorous work by \cite{Duniya:2015ths, Duniya:2016gcf}. Here we give the explicit calculations of the redshift perturbation, the density distortion, the volume distortion, and the magnification distortion, of section \ref{sec:ObsDelta_g}.

\subsection{The photon displacement}\label{appendix:PertgeoEq}
We derive the components of the perturbed vector $\delta{n}^\mu$ tangent to the perturbed geodesic $\delta{x}^\mu(\lambda)$, as discussed in section \ref{sec:TDV}. By initializing integrations at the source $S$, then integrating until the observer $O$, gives the perturbed tangent 4-vector \eqref{eq:pertgeod19}, as
\begin{align}\label{Pert:n1}
\delta{n}^\mu \Big\|^O_S  = -\Big[\bar{g}^{\mu\nu} \delta{g}_{\nu\beta}\, \bar{n}^\beta\Big]^O_S - \dfrac{1}{2}\bar{g}^{\mu\nu} \int^0_{\bar{r}_S}{ d\bar{r}\, \bar{n}^\alpha\bar{n}^\beta\partial_{\nu}\delta{g}_{\alpha\beta} },
\end{align}
where we used \eqref{eq:dr} in the integral, and that $\bar{r}(\bar{\eta}_S) = \bar{r}_S$ and $\bar{r}(\bar{\eta}_O) = 0$. Then given that $\delta{n}^\mu \equiv \delta{n}^\mu \|^S_O = -\delta{n}^\mu \|^O_S$, we obtain
\begin{eqnarray}\label{Pert:n0}
\delta{n}^0 &=& \delta{g}_{0\beta}\, \bar{n}^\beta - \dfrac{1}{2} \int^0_{\bar{r}_S}{ d\bar{r}\, \delta{g}'_{\alpha\beta} \bar{n}^\alpha\bar{n}^\beta }, \\ \label{Pert:ni}
\delta{n}^i &=& -\bar{g}^{ij} \delta{g}_{j\beta}\, \bar{n}^\beta + \dfrac{1}{2}\bar{g}^{ij} \int^0_{\bar{r}_S}{ d\bar{r}\, \partial_j (\delta{g}_{\alpha\beta}) \bar{n}^\alpha\bar{n}^\beta },
\end{eqnarray}
which give the temporal and spatial components of the perturbed tangent 4-vector. The terms outside the integrals are evaluated at $S$.

To compute the deviation 4-vector (see subsection \ref{sec:TDV}), i.e.~the 4-displacement which describes infinitesimal deviations in motion of objects away from their background world lines, we use  \eqref{pertGeod}, \eqref{Pert:n0} and \eqref{Pert:ni} as follows
\begin{align}\label{Pert:xi}
\delta{x}^i =& -\int^{\bar{r}_S}_0{d\lambda\, \left(\delta{n}^i - \bar{n}^i \delta{n}^0\right)} \nonumber\\
=& -\dfrac{1}{2}\int^{\bar{r}_S}_0{d\bar{r} \left\lbrace  \int^{\bar{r}}_{\bar{r}_S}{ d\tilde{r} \left[ \bar{g}^{ij} \partial_j(\delta{g}_{\alpha\beta}) + \delta{g}'_{\alpha\beta} \bar{n}^i\right] \bar{n}^\alpha \bar{n}^\beta } \right\rbrace  } \nonumber\\
& \quad + \int^{\bar{r}_S}_0{ d\bar{r}\, \left(\bar{g}^{ij} \delta{g}_{j\beta} + \delta{g}_{0\beta}\bar{n}^i\right)\bar{n}^\beta } \nonumber\\
=& - \dfrac{1}{2}\int^{\bar{r}_S}_0{d\bar{r} \left(\bar{r} -\bar{r}_S\right) \left[ \bar{g}^{ij} \partial_j(\delta{g}_{\alpha\beta}) + \delta{g}'_{\alpha\beta} \bar{n}^i\right] \bar{n}^\alpha \bar{n}^\beta } \nonumber\\
& \quad + \int^{\bar{r}_S}_0{ d\bar{r}\, \left(\bar{g}^{ij} \delta{g}_{j\beta} + \delta{g}_{0\beta}\bar{n}^i\right)\bar{n}^\beta }, 
\end{align}
where we obtain the last line by integrating the inner integral in the second line by parts once -- and neglecting surface terms. Equation \eqref{Pert:xi} thus gives the deviation 4-vector, incurred on a geodesic, in an inhomogeneous Universe.

\subsection{The Redshift Perturbation}\label{appendix:RSpert}
The distortion \eqref{pertRedshift} in the cosmological redshift \eqref{redshift}, evaluates as follows
\begin{align}\label{delz}
\dfrac{\delta{z}}{1+\bar{z}} &= \Big[{\bf n}\cdot{\bf v} + \phi - \sigma'\Big]^O_S - \int^0_{\bar{r}_S}{ d\bar{r}\, (\Phi' + \Psi') }, \nonumber\\
&= -\Big[V_\parallel + \Phi + \Psi - \psi\Big]^S_O - \int^0_{\bar{r}_S}{ d\bar{r}\, (\Phi' + \Psi') },
\end{align}
where in the second line we have (hereafter) neglected the unmeasurable term $\partial_r E'$ in ${\bf n}\cdot{\bf v}$, and $V_\parallel = \bar{n}^i\partial_i V$ with $V$ given by \eqref{InvVel} -- i.e.~we have also used \eqref{InvPhi} and \eqref{InvPsi}. The intrinsic peculiar velocity potential is given by
\begin{equation}
{\bf n}\cdot{\bf v} = V_\parallel - \bar{n}^i\partial_i E' .
\end{equation} 
Moreover, in \eqref{delz} we used \eqref{eq:dr}, \eqref{InvPhi}, \eqref{InvPsi} and \eqref{pert_n0}, and 
\begin{eqnarray}
\delta(n^\mu u_\mu) &=& \delta(g_{\mu\nu} n^\mu u^\nu) \nonumber\\
&=& \bar{g}_{\mu\nu} \bar{n}^\mu \delta{u}^\nu + \bar{g}_{\mu\nu} \delta{n}^\mu \bar{u}^\nu + \delta{g}_{\mu\nu} \bar{n}^\mu \bar{u}^\nu.
\end{eqnarray}
We have also used that $\bar{n}^\alpha\bar{n}^\beta \delta{g}'_{\alpha\beta} = (\delta{g}_{\alpha\beta}\, \bar{n}^\alpha\bar{n}^\beta)'$, and 
\begin{align}\label{nndelg}
\delta{g}_{\alpha\beta}\, \bar{n}^\alpha\bar{n}^\beta = -2\left[\Phi +\Psi - \dfrac{dB}{d\lambda} - \left(\dfrac{d^2E}{d\lambda^2} -2\dfrac{dE'}{d\lambda} + E''\right) \right],
\end{align}
where the metric tensor $g_{\mu\nu}$ is given by the coefficients of the spacetime differentials of the metric \eqref{Conf:metric}. We used that for any scalar $X$:
\begin{equation}\label{dXdlam}
\dfrac{dX}{d\lambda} = X' + \bar{n}^i\partial_i{X},
\end{equation}
which therefore yields that
\begin{equation}\label{nBFour}
\bar{n}^i B_{\mid  i} = \dfrac{dB}{d\lambda} - B',
\end{equation}
and (given that $E_{ij} = E_{\mid  ij} - \frac{1}{3}\delta_{ij} \nabla^2 E$) we have
\begin{eqnarray}\label{nnEFour}
\bar{n}^i \bar{n}^j E_{ij} &=& \bar{n}^i \bar{n}^j E_{\mid  ij} - \dfrac{1}{3}\nabla^2 E,\nonumber\\
&=& \left[\dfrac{d^2E}{d\lambda^2} - 2\dfrac{dE'}{d\lambda} + E''\right] - \dfrac{1}{3}\nabla^2 E,
\end{eqnarray}
where $\delta_{ij}\bar{n}^i\bar{n}^j = \bar{n}^i\bar{n}_i = 1$. In Fourier space the total Laplacian will transform as $\nabla^2 \to -k^2$. Moreover, we note that partial derivatives are commutative.

\subsection{The Radial and Angular Perturbations}\label{appendix:RThVrperts}
Here we compute the perturbations in the comoving radial distance $r$ and the polar zenith and azimuthal angles, $\theta$ and $\vartheta$,  respectively. Then if we consider \eqref{polCartTrans}--\eqref{VectCompts} and \eqref{Pert:xi}, we get that
\begin{eqnarray}\label{del:r}
\delta{r} &=& -\bar{n}_i \delta{x}^i \nonumber\\
&=& \dfrac{1}{2}\int^{\bar{r}_S}_0{d\bar{r} \left(\bar{r} -\bar{r}_S\right) \bar{n}^\nu \partial_\nu(\delta{g}_{\alpha\beta}) \bar{n}^\alpha \bar{n}^\beta }  -\int^{\bar{r}_S}_0{ d\bar{r}\, \delta{g}_{\alpha\beta} \bar{n}^\alpha\bar{n}^\beta } \nonumber\\
&=& -\dfrac{1}{2}\int^{\bar{r}_S}_0{ d\bar{r}\, \delta{g}_{\alpha\beta} \bar{n}^\alpha\bar{n}^\beta },
\end{eqnarray}
where given \eqref{eq:dr} we use that $d/d\lambda = \bar{n}^\nu \partial_\nu = -d/d\bar{r}$ (i.e.~to lowest order). We have integrated the second integral in the first line, by parts once -- and applied the stationary condition to get vanishing surface terms.

Similarly, given \eqref{polCartTrans}--\eqref{VectCompts}, it straightly follows that
\begin{align}\label{del:theta}
\bar{r}_S\,\delta{\theta} =\;& e_{\theta i}\,\delta{x}^i ,\nonumber\\
=\;& - \dfrac{1}{2}\, \int^{\bar{r}_S}_0{ d\bar{r} (\bar{r} - \bar{r}_S) e^j_{\theta}\,\partial_j \left(\delta{g}_{\alpha\beta} \right)\bar{n}^\alpha \bar{n}^\beta} + \int^{\bar{r}_S}_0{ d\bar{r} \delta{g}_{j\beta} \, e^j_{\theta}\, \bar{n}^\beta }, 
\end{align}
where $e^i_{\theta}\, \bar{n}_i =0$, by orthogonality. Moreover, \eqref{polCartTrans}--\eqref{VectCompts}, yield
\begin{align}\label{del:vartheta}
\bar{r}_S\,\sin{\theta}\,\delta{\vartheta} =\;& e_{\vartheta i}\,\delta{x}^i , \nonumber\\
=\;& - \dfrac{1}{2}\, \int^{\bar{r}_S}_0{ d\bar{r} (\bar{r} - \bar{r}_S) e^j_{\vartheta}\,\partial_j \left(\delta{g}_{\alpha\beta} \right) \bar{n}^\alpha \bar{n}^\beta } + \int^{\bar{r}_S}_0{ d\bar{r} \delta{g}_{j\beta} \, e^j_{\vartheta}\, \bar{n}^\beta },
\end{align}
where also, $e^i_{\vartheta}\, \bar{n}_i =0$. Thus \eqref{del:r}--\eqref{del:vartheta} give the explicit expressions for the perturbations $\delta{r}$, $\delta{\theta}$ and $\delta{\vartheta}$.

\subsection{The Volume Distortion}\label{appendix:Volpert}
Similarly, in the polar coordinates the various components of the $3$-gradient, are given by
\begin{equation}\label{nabCompts}
\partial_r = -\bar{n}^i\partial_i,~~\dfrac{1}{r}\, \partial_{\theta} = {e}^i_{\theta}\partial_i,~~\dfrac{1}{r\,\sin{\theta}}\,\partial_{\vartheta} = {e}^i_{\vartheta}\partial_i,
\end{equation}
where we set ${e}^i_r =-\bar{n}^i$. Hence we have that \cite{Bonvin:2011bg}
\begin{align}\label{e_theta}
e^j_{\theta} \partial_j \left(\delta{g}_{\alpha\beta} \right) \, \bar{n}^\alpha \bar{n}^\beta &= \dfrac{1}{\bar{r}}\, \left[ \partial_{\theta}  (\delta{g}_{\alpha\beta} \, \bar{n}^\alpha \bar{n}^\beta ) - \delta{g}_{\alpha\beta}\, \partial_{\theta} (\bar{n}^\alpha\bar{n}^\beta ) \right] \nonumber\\
&= \dfrac{1}{\bar{r}}\, \left[ \partial_{\theta} ( \delta{g}_{\alpha\beta} \, \bar{n}^\alpha \bar{n}^\beta ) + 2\, \delta{g}_{\alpha j} \,\bar{n}^\alpha e^j_{\theta} \right], 
\end{align} 
where we used that $\partial_{\theta} \bar{n}^\alpha = \delta^\alpha\/_i \partial_{\theta} \bar{n}^i = -\delta^\alpha\/_i e^i_{\theta}$. Similarly, given \eqref{nabCompts}, we have
\begin{eqnarray}\label{e_vartheta}
e^j_{\vartheta} \partial_j \left(\delta{g}_{\alpha\beta} \right) \, \bar{n}^\alpha \bar{n}^\beta &=& \dfrac{1}{\bar{r}\,\sin{\theta}}\, \left[ \partial_{\vartheta} ( \delta{g}_{\alpha\beta} \, \bar{n}^\alpha \bar{n}^\beta ) + 2 \delta{g}_{\alpha j} \,\bar{n}^\alpha e^j_{\vartheta} \sin{\theta} \right],
\end{eqnarray}
where $\partial_{\vartheta} \bar{n}^\alpha = -\delta^\alpha\/_i e^i_{\vartheta}\, \sin{\theta}$. Then given \eqref{del:theta}, \eqref{del:vartheta}, \eqref{e_theta} and \eqref{e_vartheta}, we get the perturbation in the volume due to the angular perturbations, given by \cite{Bonvin:2011bg}
\begin{eqnarray}\label{defVolOm}
\delta_{\Omega} &\equiv & \left(\cot{\theta_O} + \partial_{\theta}\right) \delta{\theta} + \partial_{\vartheta}\delta{\vartheta}, \\ \label{VolOm}
&=& \dfrac{1}{2} \int^{\bar{r}_S}_0 d\bar{r} \dfrac{1}{\bar{r}} \Big[ \left(\cot{\theta_O} + \partial_{\theta}\right) (\delta{g}_{i\beta} e^i_{\theta} \bar{n}^\beta) + \dfrac{\partial_{\theta}}{\sin{\theta}} \left(\delta{g}_{i\beta} e^i_{\vartheta} \bar{n}^\beta \right) \Big] \nonumber\\
&& -\; \dfrac{1}{2} \int^{\bar{r}_S}_0{ d\bar{r} \dfrac{\left(\bar{r}_S - \bar{r}\right)}{\bar{r}_S \bar{r}} \nabla^2_{\Omega}  (\delta{g}_{\alpha\beta} \bar{n}^\alpha \bar{n}^\beta) }  ,
\end{eqnarray}
where the angular part of the Laplacian is given by
\begin{equation}
\nabla^2_{\Omega} \;\equiv\; \left(\cot{\theta_O} + \partial_{\theta}\right) \partial_{\theta} + \dfrac{1}{\sin{\theta}} \partial^2_{\vartheta}.
\end{equation}
Furthermore, we compute the following terms -- i.e.~given \eqref{dXdlam} and \eqref{nabCompts},
\begin{eqnarray}\label{eTheta}
\delta{g}_{\alpha j} \,\bar{n}^\alpha e^j_{\theta} &=& \dfrac{\partial_{\theta} B}{\bar{r}} + 2\bar{n}^ie^j_{\theta} E_{\mid  ij} \nonumber\\
&=& \dfrac{\partial_{\theta}B}{\bar{r}} + \dfrac{2}{\bar{r}} \partial_{\theta}\left[\dfrac{dE}{d\lambda} - E'\right] .
\end{eqnarray}
Then in a similar manner, we obtain that
\begin{equation}\label{eVarTheta}
\delta{g}_{\alpha j} \,\bar{n}^\alpha e^j_{\vartheta} \;=\; \dfrac{\partial_{\vartheta}B}{\bar{r}\sin{\theta}} + \dfrac{2}{\bar{r}\sin{\theta}} \partial_{\vartheta}\left[\dfrac{dE}{d\lambda} - E'\right].
\end{equation}

By using \eqref{VolOm}, \eqref{eTheta} and \eqref{eVarTheta}, the angular volume perturbation -- i.e.~the angular part of the volume density perturbation -- thus becomes
\begin{eqnarray} \label{delOm}
\delta_{\Omega} &=& -\int^{\bar{r}_S}_0{ d\bar{r} \dfrac{\left(\bar{r}_S - \bar{r}\right)}{\bar{r}_S \bar{r}} \nabla^2_{\Omega} \left(\Phi + \Psi\right)} \nonumber\\
&& +\; \int^{\bar{r}_S}_0{ d\bar{r} \dfrac{1}{\bar{r}} \nabla^2_{\Omega}\left[\dfrac{B}{\bar{r}} +\dfrac{2}{\bar{r}} \left(\dfrac{dE}{d\lambda} -E'\right) \right] }\nonumber\\
&& -\int^{\bar{r}_S}_0{ d\bar{r} \dfrac{(\bar{r}_S -\bar{r})}{\bar{r}_S \bar{r}} \nabla^2_{\Omega}\left[\dfrac{dB}{d\lambda} + \left(\dfrac{d^2E}{d\lambda^2} -2\dfrac{dE'}{d\lambda}\right) \right] } \nonumber\\
&=& -\int^{\bar{r}_S}_0{ d\bar{r} \left(\bar{r}_S - \bar{r}\right) \dfrac{\bar{r}}{\bar{r}_S} \nabla^2_\perp \left(\Phi + \Psi\right)} - \left[\nabla^2_\perp E\right]^S_O,
\end{eqnarray}
where in the second line we have done integration by parts once, and set surface terms to vanish. We have also used \eqref{eq:dr}, and 
\begin{equation}\label{nablaPerp}
\nabla^2_{\perp} \equiv \dfrac{1}{\bar{r}^2} \nabla^2_{\Omega} = \nabla^2 - \partial^2_r - \dfrac{2}{\bar{r}} \partial_r ,
\end{equation}
which is the image-plane Laplacian, i.e.~in the plane of the source, perpendicular to the line of sight. Thus the quantity $\nabla^2_{\perp}E$, is transverse to the photon geodesic, with
\begin{align} \label{nabPerpE}
\nabla^2_{\perp}E = \nabla^2 E - \left(\dfrac{d^2E}{d\lambda^2} -2\dfrac{dE'}{d\lambda} + E''\right) + \dfrac{2}{\bar{r}} \left(\dfrac{dE}{d\lambda} -E'\right), 
\end{align}
i.e.~in Fourier space. Here we have used \eqref{dXdlam}, \eqref{nabCompts} and \eqref{nablaPerp}. 

Then the volume density contrast \eqref{delta_V3}, is given by
\begin{align}\label{dV1}
\check{\delta}_{\cal V} =\;&  -3D  - V_\parallel  - \dfrac{d\delta{r}}{d\lambda}  +  \dfrac{a}{{\cal H}} \dfrac{d\delta{z}}{d\lambda}  + 2\dfrac{\delta{r}}{\bar{r}_S} + \delta_{\Omega} + a\left(4 - \dfrac{2}{\bar{r}_S {\cal H}} - \dfrac{{\cal H}'}{{\cal H}^2} \right) \delta{z},\\ \label{dV2}
 =\;& \int^{\bar{r}_S}_0{ d\bar{r}\left[\dfrac{2}{\bar{r}_S} - \left(\bar{r}_S - \bar{r}\right) \dfrac{\bar{r}}{\bar{r}_S} \nabla^2_\perp\right] \left(\Phi + \Psi\right) }  + 3\int^{\bar{r}_S}_0{ d\bar{r} \left(\Phi' + \Psi'\right) } \nonumber\\
& +\; \left(\dfrac{2}{\bar{r}_S {\cal H}} + \dfrac{{\cal H}'}{{\cal H}^2} \right) \Big[ {\cal H}E - {\cal H}B  + \Phi + V_\parallel  - \int^{\bar{r}_S}_0{d\bar{r} \left(\Phi' + \Psi' \right) } \Big] \nonumber\\
& +\; \left(E'' +\dfrac{2}{\bar{r}_S}E'\right) + \dfrac{2}{\bar{r}_S} \left(B - 2E'\right) - \dfrac{dB}{d\lambda} - 4V_\parallel - 2\left(\Phi + \Psi\right) \nonumber\\
 & -\; \dfrac{1}{{\cal H}} \left[\dfrac{d}{d\lambda} \left(-\psi + V_\parallel \right) + \dfrac{d\Psi}{d\lambda} -\Psi' - \partial_r\Phi \right], 
\end{align}
where given \eqref{eq:dr}, \eqref{delz}, \eqref{nndelg} and \eqref{del:r}, we used
\begin{eqnarray}
\dfrac{a}{{\cal H}} \dfrac{d\delta{z}}{d\lambda} &=& \Phi + \Psi -\psi + V_\parallel  - \int^{\bar{r}_S}_0{d\bar{r} \left(\Phi' + \Psi'\right) } \nonumber\\
&-& \dfrac{1}{{\cal H}} \left[\dfrac{d}{d\lambda} \left(-\psi + V_\parallel\right) - \partial_r \left(\Phi + \Psi\right)\right], 
\end{eqnarray}
and the perturbation in the comoving radial distance as
\begin{align}\label{2dr_r}
2\dfrac{\delta{r}}{\bar{r}_S} &= \dfrac{2}{\bar{r}_S} \int^{\bar{r}_S}_0{d\bar{r} \left(\Phi + \Psi\right)} + \dfrac{2}{\bar{r}_S} \left[B + \left(\dfrac{dE}{d\lambda} - 2E'\right) \right] ,
\end{align} 
with the total derivative given by
\begin{equation}
\dfrac{d\delta{r}}{d\lambda} = -\left[\Phi +\Psi - B - \left(\dfrac{dE}{d\lambda} - 2E'\right) \right] .
\end{equation}
Then from \eqref{dV2}, we make the following simplification
\begin{align}
\Big(E'' &+\dfrac{2}{\bar{r}_S}E'\Big) + \dfrac{2}{\bar{r}_S} \left(B - 2E'\right) - \dfrac{dB}{d\lambda} +  {\cal H}\left(\dfrac{2}{\bar{r}_S {\cal H}} + \dfrac{{\cal H}'}{{\cal H}^2} \right) \left[E - B\right] = \nonumber\\
&\hspace{2cm} -\left(\dfrac{dB}{d\lambda} + \dfrac{{\cal H}'}{{\cal H}}B \right) + \left(\dfrac{dE'}{d\lambda} + \dfrac{{\cal H}'}{{\cal H}}E' \right) ,\nonumber 
\end{align}
and given \eqref{InvPsi}, we have 
\begin{eqnarray}
\dfrac{1}{{\cal H}} \dfrac{d\Psi}{d\lambda} = \dfrac{1}{{\cal H}} \dfrac{d}{d\lambda} \left(D + \dfrac{1}{3}\nabla^2E \right) + \dfrac{dE'}{d\lambda} + \dfrac{{\cal H}'}{{\cal H}}E'  - \left(\dfrac{dB}{d\lambda} + \dfrac{{\cal H}'}{{\cal H}} B \right) .
\end{eqnarray}

Finally we have, the total volume density perturbation given by
\begin{eqnarray}\label{delVol1}
\check{\delta}_{\cal V} &=& -\int^{\bar{r}_S}_0{ d\bar{r} \dfrac{(\bar{r}_S - \bar{r})}{\bar{r}_S\bar{r}} \nabla_{\Omega} \left(\Phi + \Psi\right) } - 4V_\parallel - 2\left(\Phi + \Psi\right) \nonumber\\
 && +\; 3\int^{\bar{r}_S}_0{ d\bar{r} \left(\Phi' + \Psi'\right) } + \dfrac{2}{\bar{r}_S}\int^{\bar{r}_S}_0{ d\bar{r} \left(\Phi + \Psi\right) } \nonumber\\
 &&  +\; \left(\dfrac{2}{\bar{r}_S {\cal H}} + \dfrac{{\cal H}'}{{\cal H}^2} \right) \left[ \Phi + V_\parallel  - \int^{\bar{r}_S}_0{d\bar{r} \left(\Phi' + \Psi' \right) } \right] \nonumber\\
&& +\; \dfrac{1}{{\cal H}} \left[\Phi' + \partial_r\Psi -\dfrac{dV_\parallel}{d\lambda}  \right] .
 \end{eqnarray}

\subsection{The Magnification Perturbation}
Consider the area density \eqref{A:defn} -- which is transverse to the photon geodesic. Then the only non-vanishing terms to yield
\begin{align}\label{DefnA1}
\mathcal{A} &\;=\; a^{-2}\sqrt{-\tilde{g}} \left[1 + \dfrac{\delta{u}^0}{\bar u^0} + \dfrac{\delta{\ell}^l}{\bar{\ell}^l}\right] \nonumber\\
& \hspace{2cm} \times \epsilon_{ijk}\,\bar{\ell}^i\, \dfrac{\partial\tilde{x}^j}{\partial\theta_S} \dfrac{\partial\tilde{x}^k}{\partial\vartheta_S} \left\| \left. \dfrac{\partial(\theta_S,\vartheta_S)}{\partial(\theta_O,\vartheta_O)} \right\| \right. ,\\ \label{DefnA2}
&\;=\; a^2 \bar{r}^2\sin{\theta_S} \Big[1 -3D -\phi + \bar{n}^i B_{\mid  i} \nonumber\\
& \hspace{2cm} - \dfrac{1}{2}\delta{g}_{\mu\nu}\bar{n}^\alpha\bar{n}^\beta\Big] \left\| \left. \dfrac{\partial(\theta_S,\vartheta_S)}{\partial(\theta_O,\vartheta_O)} \right\| \right., \\ \label{DefnA3}
&\;=\; a^2\bar{r}^2\sin{\theta_O}\left[1 -3D -\phi + \bar{n}^i B_{\mid  i} - \dfrac{1}{2}\delta{g}_{\mu\nu}\bar{n}^\alpha\bar{n}^\beta \right. \nonumber\\
&  \hspace{1.5cm} \left. +\; 2\dfrac{\delta{r}}{\bar{r}} + \left(\cot{\theta_O} +\partial_\theta\right)\delta{\theta} +\partial_\vartheta\delta{\vartheta}\right],
\end{align}  
where $\sqrt{-\tilde{g}}$ is given by \eqref{eq:39}, with $\bar{\tilde u}^\mu = a^{-1}\delta^\mu\/_0$ and $\bar{\tilde u}_\mu = -a\delta^0\/_\mu$ as given by \eqref{vels}. The determinant of the transformation matrix is given by $|J|= 1 +\partial_{\theta}\delta{\theta} + \partial_{\vartheta}\delta{\vartheta}$ \eqref{eq:37}, and for small $\delta\theta$ we have $\sin{\theta_S} = (1 + \delta{\theta}\cot{\theta_O}) \sin{\theta_O}$. Also, \eqref{ell:defn} becomes
\begin{equation}
\tilde{\ell}^\nu = a^{-1} \left(u^\nu + \dfrac{n^\nu}{n^\alpha u_\alpha} \right) =  a^{-1}\ell^\nu .
\end{equation}
Given \eqref{n:Conf} and \eqref{vels} we have $\bar{\ell}^0 = 0$ and $\bar{\tilde \ell}^i = a^{-1}\bar{\ell}^i = -a^{-1}\bar{n}^i / \bar{n}^0$. Then,
\begin{eqnarray}
\dfrac{\delta{\ell}^i}{\bar{\ell}^i} &=& -\bar{n}_i v^{\mid  i} + \bar{n}_i\delta{n}^i - \delta{n}^0 + \delta{u}^0 - \bar{\ell}^i\delta{u}_i, \\
&=& \bar{n}^i B_{\mid  i} - \phi - \dfrac{1}{2} \delta{g}_{\alpha\beta} \bar{n}^\alpha\bar{n}^\beta,
\end{eqnarray}
where in the first line we used the identity $\bar{n}_i = 1/\bar{n}^i$, and the second line comes by combining \eqref{Pert:n0} and \eqref{Pert:ni} and integrating once: with the integrals as indefinite integrals (i.e.~with the limits dropped). By using \eqref{InvPsi} and \eqref{dXdlam}, we get 
\begin{align}\label{delpsi1}
\delta_\phi &\equiv  -3D -\phi + \bar{n}^i B_{\mid  i} - \dfrac{1}{2}\delta{g}_{\mu\nu}\bar{n}^\alpha\bar{n}^\beta ,\\ \label{delpsi2}
&= -2\Psi -E +2{\cal H}\sigma - \sigma' - B' - \left(\dfrac{d^2E}{d\lambda^2} -2\dfrac{dE'}{d\lambda}\right) .
\end{align} 
Then given \eqref{delOm}, \eqref{nabPerpE}, \eqref{2dr_r}, \eqref{DefnA3} and \eqref{delpsi2}, we obtain
\begin{eqnarray}\label{matA1}
\mathcal{A} &=& a^2\bar{r}^2\sin{\theta_O} \left\{1 -2\Psi + \int^{\bar{r}_S}_0{ d\bar{r}\left[\dfrac{2}{\bar{r}_S} + \left(\bar{r} -\bar{r}_S\right) \dfrac{\bar{r}}{\bar{r}_S} \nabla^2_\perp \right]\left(\Phi +\Psi\right) } \right.\nonumber\\
&& \hspace{2cm} \left. + 2{\cal H}\left(1 - \dfrac{1}{\bar{r}_S{\cal H}} \right) \sigma \right\} .
\end{eqnarray}
By taking a gauge transformation, we get the redshift-space perturbation
\begin{eqnarray}\label{dmatA1}
\check{\delta}_{\cal A} &=& \delta_{\cal A} - \dfrac{d\ln\bar{\cal A}}{d\bar{z}} \delta{z}, \\ \label{dmatA2}
&=& -2\Psi + \int^{\bar{r}_S}_0{ d\bar{r}\left[\dfrac{2}{\bar{r}_S} + \left(\bar{r} -\bar{r}_S\right) \dfrac{\bar{r}}{\bar{r}_S} \nabla^2_\perp \right]\left(\Phi +\Psi\right) } \nonumber\\
&& +\; 2\left(1 - \dfrac{1}{\bar{r}_S{\cal H}} \right) \left[{\cal H}\sigma + a\,\delta{z}\right],
\end{eqnarray} 
where $\mathcal{\bar A} = a^2\bar{r}^2\sin{\theta_O}$, and 
\begin{equation}
\dfrac{d\ln{\mathcal{\bar A}}}{d\bar{z}} \;=\; 2a\left(\dfrac{1}{\bar{r}_S{\cal H}} - 1\right).
\end{equation}
Thus given \eqref{delz}, we get
\begin{eqnarray}\label{dmag1}
\mu^{-1} &=& 1 - \check{\delta}_{\cal M} = 1 + \check{\delta}_{\cal A},\\ \label{dmag2}
&=& 1 - 2\Psi + \int^{\bar{r}_S}_0{ d\bar{r}\left[\dfrac{2}{\bar{r}_S} + \left(\bar{r} -\bar{r}_S\right) \dfrac{\bar{r}}{\bar{r}_S} \nabla^2_\perp \left(\Phi +\Psi\right) \right] } \nonumber\\
&& -\; 2\left(1 - \dfrac{1}{\bar{r}_S{\cal H}} \right) \left[\Phi +V_\parallel - \int^{\bar{r}_S}_0{ d\bar{r} \left(\Phi' + \Psi'\right) }\right] .
\end{eqnarray}
The magnification contrast $\check{\delta}_{\cal M}$ is hence given by \eqref{dmag1} -- which measures the magnification distortion (up to the magnification bias \eqref{b_M}) in the observed overdensity.

\end{appendices}




\end{document}